\documentclass[a4paper,12pt]{article}
\usepackage[T1]{fontenc}
\usepackage[utf8]{inputenc}
\usepackage[affil-it]{authblk}
\usepackage{tikz}
\usetikzlibrary{decorations.text,calc,arrows.meta}
\usepackage{slashed}
\usepackage{amssymb, amsmath, amsfonts}
\usepackage{xcolor}
\usepackage{a4wide}
\usepackage{mathpazo}
\usepackage{mathrsfs} 
\usepackage{calligra}
\usepackage{hyperref}
\hypersetup{
    colorlinks,%
    citecolor=blue,%
    filecolor=green,%
    linkcolor=green!50!black,%
    urlcolor=blue
}
\usepackage{txfonts} 
\usepackage[numbers,sort&compress,merge]{natbib}
\usepackage{graphicx}
\usepackage{subfig}
\usepackage[toc,page]{appendix}
\usepackage{cancel}
\DeclareMathAlphabet{\mathpzc}{OT1}{pzc}{m}{it}
\DeclareFontShape{T1}{calligra}{m}{n}{<->s*[1.8]callig15}{}
\newtheorem{theorem}{Theorem}[section]
\newtheorem{proposition}{Proposition}[section]
\newtheorem{lemma}{Lemma}[section]
\newtheorem{mydef}{Definition}[section]
\newtheorem{remark}{Remark}[section]
\newcommand{\scalarProd}[2]{ \left\langle {#1} , {#2} \right\rangle } 
\newcommand{\norm}[1]{\left| \left| {#1} \right| \right|}

\newcommand{\sA}{\mathscr{A}}
\newcommand{\fA}{\mathsf{A}}

\newcommand{\sB}{\mathscr{B}}
\newcommand{\fB}{\mathsf{B}}

\newcommand{\sC}{\mathscr{C}}
\newcommand{\fC}{\mathsf{C}} 
\newcommand{\C}{\mathbb{C}}

\newcommand{\sD}{\mathscr{D}}

\newcommand{\sF}{ \mathscr{F} }

\newcommand{\sH}{\mathscr{H}}

\newcommand{\Iw}{I_{\omega} }
\newcommand{\IwV}{I_{\omega, \scriptscriptstyle{V}} }

\newcommand{\sK}{\mathscr{K}}
\newcommand{\ck}{\mathpzc{k}}

\newcommand{\sM}{\mathscr{M}}
\newcommand{\fU}{\mathsf{U}}
\newcommand{\sV}{\mathscr{V}}
\newcommand{\sW}{\mathscr{W}}

\newcommand{\DeltaVrep}{	\Delta_{ \omega, \scriptscriptstyle{V} }	}

\newcommand{\DeltaBvac}{	\Delta_{ 0, \scriptscriptstyle{B'} }	}

\newcommand{\deltaBvac}{	\delta_{ 0, \scriptscriptstyle{B'} }	}
\newcommand{\deltaVvac}{	\delta_{ 0, \scriptscriptstyle{V} }	}

\newcommand{\deltaVrep}{	\delta_{ \omega, \scriptscriptstyle{V} }	}
\newcommand{\deltaVLargerep}{	\delta_{ \omega, \scriptscriptstyle{\hat{V}} }	}

\newcommand{\JVrep}{	J_{ \omega, \scriptscriptstyle{V} }	}

\newcommand{\jVrep}{	j_{ \omega, \scriptscriptstyle{V} }	}

\newcommand{\TomitaVrep}{	S_{ \omega, \scriptscriptstyle{V} }	}

\newcommand{\tomitaVvac}{	s_{ 0, \scriptscriptstyle{V} }	}
\newcommand{\tomitaVrep}{	s_{ \omega, \scriptscriptstyle{V} }	}

\newcommand{\fT}{\mathsf{T}}

\newcommand{\LtwoCauchy}{L^{2} (\mathscr{S\!M} \! \upharpoonright \! \mathscr{C})}

\newcommand{\rN}{\mathrm{N}}
\newcommand{\N}{\mathbb{N}}
\newcommand{\R}{\mathbb{R}}
\newcommand{\sR}{\mathscr{R}}

\newcommand{\spinorBun}{ \mathscr{S\!M}}
\newcommand{\spinorBunCauchy}{ \mathscr{S\!M} \! \upharpoonright \! \mathscr{C} }

\newcommand{\ri}{ \mathrm{i} }

\newcommand{\clo}[1]{ \mathrm{clo} ({#1}) } 
\newcommand{\supp}[1]{ \mathrm{supp} ({#1}) }

\newcommand{\fbracket}[1]{ \left( {#1} \right) } 
\newcommand{\sbracket}[1]{ \left\{ {#1} \right\} }
\newcommand{\tbracket}[1]{ \left[ {#1} \right] }
\newcommand{\one}{\mathbb{1}} 

\newcommand{\ie}{i.e.,~}
\newcommand{\dist}{\mathrm{dist}}

\newcommand{\sh}{ {\footnotesize \textrm{\calligra{h}}}~ }

\newcommand{\PA}{	P_{ \scriptscriptstyle{A} }	}
\newcommand{\PB}{	P_{ \scriptscriptstyle{B} }	}
\newcommand{\PV}{	P_{ \scriptscriptstyle{V} }	}

\newcommand{\PVlarge}{	P_{ \hat{\scriptscriptstyle{V} } }	}

\newcommand{\SigmaB}{	\Sigma_{ 0, \scriptscriptstyle{B}' }}

\newcommand{\SigmaOmegaV}{     \Sigma_{ \omega, \scriptscriptstyle{V} }	}
\newcommand{\SigmaOmegaVLarge}{     \Sigma_{ \omega, \scriptscriptstyle{\hat{V}} }	}

\newcommand{\ms}{\scriptscriptstyle}

\newcommand{\Tr}{\mathrm{Tr}}

\title{Relative entanglement entropy for widely separated regions in curved spacetime}
\author[1]{Stefan Hollands\thanks{stefan.hollands@uni-leipzig.de}}
\author[1]{Onirban Islam\thanks{onirban.islam@itp.uni-leipzig.de}}
\author[2]{Ko Sanders\thanks{jacobus.sanders@dcu.ie}}
\affil[1]{Institut  f\"{u}r  Theoretische  Physik, Universit\"{a}t  Leipzig, \protect \\
    Br\"{u}derstra\ss{}e  16,  D-04103  Leipzig, Germany}
\affil[2]{Dublin City University, School of Mathematical Sciences, \protect \\
    Glasnevin, Dublin 9, Ireland}
\date{6 November 2017}

\begin{document}
    
    \maketitle

    \begin{abstract}
	We give an upper bound of the relative entanglement entropy of the ground state 
    of a massive Dirac-Majorana field across two widely separated regions $A$ and $B$ 
    in a static slice of an ultrastatic Lorentzian spacetime. 
    Our bound decays exponentially in $ \dist (A, B) $, at a rate set by the Compton wavelength and the spatial scalar curvature. 
    The physical interpretation our result is that, 
    on a manifold with positive spatial scalar curvature, 
    one cannot use the entanglement of the vacuum state to teleport one classical bit 
    from $A$ to $B$ if their distance is of the order of the maximum of the curvature 
    radius and the Compton wave length or greater.  
    \end{abstract}

    \section{Introduction}
    \label{sec: introduction}
    In quantum field theory, a global state is called entangled across a pair of spacelike 
	separated subsystems if its restriction to the bipartite system is not in 
	the weak $*$-closure of the convex hull of normal product states~\cite{CliftonHalvorson_HPMP2001}. 
	A systematic way of quantifying entanglement is to introduce entanglement measures~\cite{Vedral_PRL1997}. 
	These are state functionals which, at a bare minimum, vanish on separable states and are
	non-increasing under local operations and  classical communications. 
	(For more details on this in the quantum field theory setting, see~\cite{HollandsSanders_2017} and references therein.) 
	It can be argued that the ``relative entanglement entropy'' is a particularly good entanglement measure in quantum field theory.
	For instance it has been shown~\cite[Thm. 11]{HollandsSanders_2017} that 
	the relative entanglement entropy $E$ of the ground state of Dirac-Majorana field theory across two disjoint regions 
	$A$ and $B$ in a static slice of a $D$-dimensional ultrastatic spacetime has an upper bound of the form 
	$ E \le \mathrm{cst}.~| \log (m d) | d^{ - (D - 2)} | \partial A | $ when $ d := \dist (A, B) \to 0  $ 
	and $m$ is the mass of Dirac-Majorana field.

	The opposite limit, \ie when the subsystems are far apart from each other is  addressed in this article. 
	We find that the relative entanglement across $A$ and $B$ is bounded above by
	$ E \le \mathrm{cst}. \, \exp(-M d / 2) $, when $ d \to \infty $, 
	and where $ M := \inf( m^2 + R / 4 )^{1/2} $ is an effective mass  
	related to the Dirac mass $m$ and the scalar curvature $R$ of the manifold. The precise statement is given below in Thm.~\ref{thm: REE_Dirac}.

	It is known that one needs about $\log 2$ of entanglement to teleport one classical bit of information from $A$ to $B$, 
    since this is the entanglement of one Bell-pair. 
    The number of Bell pairs (one partner in $A$, the other in $B$) that one can distill from the vacuum across $A$ and $B$ 
    is not bigger then $E$ by a general theorem~\cite{Donald_PLA1999}. 
    Thus, the physical interpretation our result is that, on a manifold with positive spatial scalar curvature $R$, 
    one cannot use the entanglement of the vacuum state to teleport one classical bit if the distance exceeds in order 
    of the magnitude the maximum of the curvature radius and the Compton wave length.

	It would be interesting to connect our work to results in the physics literature on multi-region entanglement 
    such as \cite{Calabrese_JStatMech2009, CasiniHuerta_JPA2009}, or to holographic approaches, see e.g. \cite{RangamaniTakayanagi_Springer2017}.

    \section{Dirac quantum field theory on curved spacetime}
    \label{sec: quasifree}
    We begin with a very brief review of  
    the quantization of a classical, linear Dirac-Majorana field on a globally hyperbolic 
    $D$-dimensional   simply connected spin-spacetime $ (\mathscr{M}, g) $
    (for more details, see e.g.~\cite{AntoniHollands_CMP2006, baer_springer2012}). 
    Because we want (pseudo-)Majorana spinors to exist we must restrict the spacetime dimension to the values
    $ D = 3, 4, 8, 9, 10 ~\mathrm{mod}~ 8$~\cite{KugoTownsend_NuclPhysB1983, Trautman_JGP2008}. 
    In fact, because we will need to define a time-reversal operator below, we must further restrict to dimensions 
    $ D = 4, 8, 9, 10 ~\mathrm{mod}~ 8$.
    For the purpose of this work, we need the algebras of observables associated with two regions 
    $ A, B $ in a Cauchy surface $\sC \subset \sM $.  
    These are defined most elegantly in the general framework of Araki~\cite{Araki_RIMS1970} 
    which we briefly recall in this section.

    First, we explain our conventions regarding complex vector spaces and linear maps. 
    For any complex vector space $\sV$, the complex conjugate vector space $ \bar{\sV} $ 
    is identical as a set but equipped with the scalar multiplication 
    $ \lambda \cdot v = \bar{\lambda} v $. 
    The identity map between the sets $\sV$ and $ \bar{\sV} $ is denoted by $ v \mapsto \bar{v} $ and is anti-linear.
    If $\mathsf{E}$ is a linear endomorphism from a complex vector space $\sV$ to a complex vector space $\sW$, 
    then $ \bar{\mathsf{E}} $ is the natural linear endomorphism from $ \bar{\sV} $ to   $ \bar{\sW}$. 
    $\sV^{\prime}$ denotes the dual space of complex linear functionals from $\sV$ to $\mathbb C$.

    Let $\spinorBun$ be the spinor bundle over $\sM$ for a chosen spin-structure. 
    We consider a given spinor representation $ \gamma : T \sM \to \mathrm{End} (\spinorBun) $. 
    Let $ \{ e^{0}, e^{1}, \dots, e^{D-1} \} $ be an oriented, time-oriented, orthonormal frame 
    field defined locally on $(\sM, g)$. 
    We let $ \gamma(e^{a}), a = 0,\dots, D - 1 $ be the corresponding elements of the 
    spinor representation, satisfying the usual anti-commutation relations 
    \begin{equation} \label{eq: def_Cliff_alg}
		\gamma(e^{a}) \gamma(e^{b}) + \gamma(e^{b}) \gamma(e^{a}) = 2 \eta^{ab} \mathbb{1}, 
	\quad \eta = \mathrm{diag} (1,-1,\dots, -1) \quad a,b = 0, \dots, D-1.
    \end{equation}
    Associated with the spinor representation there are three natural maps, $\fA, \fB, \fC$.  
    We note that the spinor representation $\gamma$ and the intertwiners $\fA, \fB, \fC$ 
    are covariantly constant with respect to the connection $\nabla$ in the 
    spinor bundle $\spinorBun$~\cite{KugoTownsend_NuclPhysB1983, Trautman_JGP2008}. 
    $\fC$ is an intertwiner between the representation 
    $ X \mapsto \gamma (X) $ on $\spinorBun$ and the conjugate representation $ X \mapsto \overline{\gamma(X)} $ on $\overline{\spinorBun}$, 
    so $ \fC: \spinorBun \to \overline{\spinorBun} $ and $\fC$ acts fibrewise. 
    $\fB$ is an intertwiner between the representation 
    $ X \mapsto \gamma(X) $ on $\spinorBun$ and the dual representation $ X \mapsto \gamma(X)^{\prime} $ on $\spinorBun^{\prime}$,
    so $ \fB: \spinorBun \to \spinorBun^{\prime} $ and $\fB$ acts fibrewise. 
    In each fibre, we can normalize $\fC$ such that  $ \bar{\fC} \fC = 1 $ in dimension $ D = 3,4,8,9,10 $ mod $8$, 
    to which we restrict for this reason in this article.
    It will be possible to make a consistent choice of a normalized map $\fC$ globally if $ H^{1} (\sM) $ is trivial, which we assume. 
    We furthermore consider $ \fA := \bar{\fB} \fC : \spinorBun \to \overline{\spinorBun}^{\prime} $. 
    It can be shown that the map $\fB$ can be normalized in a given fibre by a real constant in such a way that
    $\fA$ satisfies $ \bar{\fA}^{\prime} = \fA $. 
    Again, this choice can be made globally on a simply connected manifold.

    We now pick a Cauchy surface $\sC$ of $\sM$, and denote its forward directed time-like normal vector field by $e^{0}$.  
    We let  $ \sK := \LtwoCauchy $ be the space of square-integrable spinors defined on $\sC$ with respect to the 
    positive definite, non-degenerate hermitian inner product, see e.g.~\cite{Trautman_JGP2008} 
    \begin{equation} \label{eq: L2_inner_product}
		\langle k , k' \rangle := \int_{\sC} \prec \overline{ \gamma (e^{0}) k},  \fA(k') \succ \mathrm{d} \mathrm{vol}_{\sC}, 
    \end{equation}   
    where $ \prec \cdot \ , \ \cdot \succ$ is the pairing arising from the canonical pairing between $\spinorBun$ and its dual $\spinorBun^\prime$. 
    Our definitions imply that the map  $\Gamma$ on $\sK$ defined by $\Gamma k := \bar{\fC} \bar{k} $ is involutive, 
    $ \Gamma^{2} = 1 $, anti-linear, and unitary (we call such maps ``anti-unitary''). 
    We think of $\Gamma$ as Majorana (``charge'') conjugation. 
   In flat space and $D=4$, this inner product is often written as 
    ``$ \int k^{\dagger} k' ~\mathrm{d}^3 x $ '', but we will not use this notation here.

    The algebra of canonical anti-commutation relations $ \sD(\sC) $ over $\sC$ is now defined as the free unital *-algebra over $\C$ 
    generated by the symbols $\one$, $ \psi(k), \psi(k')^{*}, \, k, k' \in \sK $ modulo the relations
    \begin{equation}
		\psi (c k) = c \, \psi (k), 
		\; 
		\psi ( k' + k ) = \psi (k') + \psi (k) , 
		\;  
		\psi (k)^{*} = \psi (\Gamma k),  
		\;  
		\sbracket{ \psi (k), \psi (k')^{*} } = \scalarProd{k'}{k} \one,  
    \end{equation}
    where $c \in \mathbb C$.
    It has been shown in~\cite{Araki_RIMS1970} how to define the (unique) $C^*$-norm on $ \sD (\sC) $, whose 
    closure (denoted by the same symbol) then defines a $C^{*}$-algebra.

    The local field algebra $ \sD (V) $ corresponding to some bounded open region $ V \subset \sC $ is
    by definition, the $ C^{*} $-subalgebra generated by all elements of the form 
    $ \psi (k) $, where $ \supp{k} \subset V $. 
    The elements $ \psi (k) $ may be thought informally as the ``time-zero'' fields 
    ``$ \psi(k) = \int_{\sC} \psi (x) k (x) $'' averaged against a ``test function'' $k$. 
    The ``spacetime fields'' are obtained from the time zero fields via the Dirac equation. 
    They are not really needed here.

    The algebra $ \sD(V) $ has an obvious grading automorphism induced by $ \psi (k) \mapsto - \psi (k) $, 
    which allows us to define consistently the linear subspaces $\sD^{+/-} (V) $ of even/odd elements. The subspace 
    $\sD^+(V)$ is a norm-closed subspace preserved under multiplication, $*$-operation and hence a $C^{*}$-subalgebra.
    If $ A, B $ are disjoint regular open subsets of $\sC$ then $ \tbracket{ \sD^+(A), \sD^+ (B) } = \{ 0 \}$, 
    \ie the even subalgebras commute.

    An algebraic state is defined as a positive, normalized linear functional $ \omega : \sD(\sC) \to \C $. 
    In this investigation, we will restrict attention mostly  to 
    quasifree~\cite{BalslevVerbeure_CMP1968, Balslev_CMP1968, Araki_RIMS1970} states. 
    These are completely characterized by their ``two-point function'', which in turn is characterized uniquely by 
    a bounded operator $ P_{\omega} $ on $\sK$ satisfying~\cite[Lem. 3.2, 3.3]{Araki_RIMS1970} 
    \begin{equation}
		\omega ( \psi (k)^{*} \psi (k') ) = \scalarProd{k}{P_{\omega} \, k'}, 
		\quad 
		0 \leq P_{\omega} = P_{\omega}^{*} \leq 1, 
		\quad 
		P_{\omega} + \Gamma P_{\omega} \Gamma = 1. 
    \end{equation}
    We will often use the operator~\cite[Prop. 1]{Balslev_CMP1968} 
    \begin{equation}
		\Sigma_{\omega} = 2P_{\omega} - 1,  
    \end{equation}
    which consequently satisfies $ -1 \leq \Sigma_{\omega} \leq 1 $ 
    and $ \Gamma ( 1 - \Sigma_{\omega} ) \Gamma = 1 + \Sigma_{\omega} $. 
    The state is pure if and only if $ P_{\omega} $ is a projection operator.

    If we apply the Gelfand-Naimark-Segal (GNS) representation to a quasifree state $\omega$, we get a GNS-triple
    $ (\pi_{\omega}, \sF_{\omega}, \Omega_{\omega}) $ characterized as follows. $\pi_\omega$ is a representation of 
    $ \sD (\sC) $ on the GNS-Hilbert space $ \sF_{\omega} $. In our case, this Hilbert space is 
    the fermionic Fock space over a ``1-particle'' space $ \sH_{\omega} $. 
    $ \sH_{\omega} $ can be constructed from $\sK$ by dividing out 
    $ \mathrm{ker} (P_{\omega}) = \mathrm{ker} ( 1 + \Sigma_{\omega} ) / 2 $. 
    On the equivalence classes, we define the inner product
    \begin{equation}
		( [k], [k'] )_{\omega} := \frac{1}{2} \scalarProd{k}{ ( 1 + \Sigma_{\omega} ) k' }, 
    \end{equation}
    and we take the closure with respect to this inner product. 
    We let $ \Iw: \sK \to \sH_{\omega} $ be the canonical map that arises out of this construction. 
    There is then a unitary map $ U_{\omega} : \sH_{\omega} \to  \mathrm{ker} ( 1 + \Sigma_\omega )^\perp$, 
    defined by $ U_{\omega} \, \Iw := \sqrt{ (1 + \Sigma_{\omega}) / 2 } $. 
    In particular, $ ( \Iw k, \Iw k')_{\omega} = \scalarProd{k}{(1 + \Sigma_{\omega}) k'} / 2 $.

    The antisymmetric space $ \sF_{\omega} $ over the $1$-particle Hilbert space is 
    $ \sF_{\omega} := \C \oplus ( \oplus_{ n = 1 }^{\infty} \bigwedge^{n} \sH_{\omega} )$,  
    and $ \Omega_{\omega}$ is the cyclic and separating vector corresponding to the summand ``$\C$''. 
    The representation of $ \sD (\sC) $ is given in terms of creation and annihilation operators, 
    \begin{equation}
		\pi_{\omega} ( \psi (k) ) = \mathrm{a} ( \Iw~ k)^{*} + \mathrm{a} (\Iw~ \Gamma k).
    \end{equation}
    with $ \{ \mathrm{a} (h), \mathrm{a}^{*} (h') \} = (h, h')_{\omega} 1 $. 
    The representation allows us to define the v. Neumann algebra (of the even part) as
    \begin{equation}
		\sR_\omega(V) := \pi_{\omega} ( \sD^+ (V) )'',
    \end{equation} 
    where $''$ is the double commutant which coincides with the weak closure. 
    These strongly depend on the choice of $\omega$.

    A specific state, called $\omega_{0}$, will now be constructed on an ultrastatic spin-spacetime 
    $ \mathscr{M} = \R \times \sC $ with metric $ g = \mathrm{d} t^{2} - h_{ij} \mathrm{d} x^{i} \mathrm{d} x^{j} $, 
    where $ x^{i}, i=1,\dots,D-1 $ are local coordinates on $\sC$ and $h_{ij}$ does not depend on $t$. 
    We assume that the Riemannian manifold $(\sC, h)$ is complete. 
    Then the 1-particle Dirac Hamiltonian is $ H = \gamma (e^{0}) ( - \ri \gamma (e^{i}) \nabla_{e_{i}} + m) $, where $m>0$ is the Dirac mass. 
    Since $(\sC,h)$ is assumed to be complete, $H$ can be shown to be essentially self-adjoint~\cite{Wolf_Indiana1973,Chernoff_JFA1973}, 
    and so we may define measurable functions of $H$ via the spectral calculus.

    The state $\omega_{0}$ is obtained by choosing the operator $P_{0}$ to be the projection onto the positive 
    spectral subspace of $H$ or, alternatively speaking, by taking $ \Sigma_{0} = H / |H| $.
    Physically, $\omega_{0}$ is the ground state of a Dirac-Majorana field. The GNS-triple 
    is denoted in this case by $ (\pi_{0}, \sF_{0}, \Omega_{0}) $.
    The v. Neumann algebras associated with a region $ V \subset \sC $ and the even subalgebras
    are defined as the weak closures and are denoted as 
    \begin{equation}
		\sA(V) \equiv \sR_{0}(V) = \pi_{0} ( \sD^+ (V) )''. 
    \end{equation}
    Since $\Gamma$ is the Majorana conjugation operator, our theory corresponds physically to a Majorana fermion. 
    
    \section{Relative entanglement entropy and main result}
    \label{sec: REE}
    Let $A$ and $B$ be two open regions in a static Cauchy surface $\sC$ of an ultrastatic spacetime, 
    and let $\sA(A)$ and $\sA(B)$ be the corresponding v.  Neumann 
    algebras acting on the Hilbert space associated with the ground state representation, 
    as described in the previous section. 
    If the distance $ d $ is positive, then it is known that there exists\footnote{This isomorphism is closely related to the 
	``split-property'', 
	see~\cite{Antoni_Longo_JFA1983, DoplicherLongo_InventMath1984}.} 
    a canonical isomorphism~\cite{HollandsSanders_2017}
    \begin{equation} \label{eq: split_property}
		\sA (A) \vee \sA (B) \cong \sA(A) ~\bar{\otimes}~ \sA(B)
    \end{equation}
    from the v. Neumann algebra generated by $ \sA (A) $ and $ \sA (B) $ to the tensor product of these algebras, 
    acting now on $\sF_0 \otimes \sF_0$. 
    This isomorphism allows us to define normal {\em product states}. 
    Namely, if $ \omega_{\ms{A}} $ is a normal state on $ \sA (A) $, $ \omega_{\ms{B}}$ a normal state on $ \sA (B) $, 
    then there exists a unique state $ \omega_{\ms{A}} ~\bar{\otimes}~ \omega_{\ms{B}}$ on $\sA(A) ~\bar{\otimes}~ \sA(B)$ determined by $ \omega_{\ms{A}} ~\bar{\otimes}~ \omega_{\ms{B}} (ab) := \omega_{\ms{A}} (a) \omega_{\ms{B}} (b)$, 
    where $ a \in \sA (A), b \in \sA (B) $. (If, on the contrary, $ d :=  {\rm dist}(A,B) = 0 $, then it is known 
    (see e.g.~\cite[Thm. 2]{HollandsSanders_2017}) that such normal product states do not exist.)
    
    A state $\omega$ on $ \sA (A) ~\bar{\otimes}~ \sA (B) $ is called ``separable'' if it is of the form
    \begin{equation}
		\sigma = \sum_{n} \omega_{ {\ms{A}}, n } ~\bar{\otimes}~ \omega_{ {\ms{B}}, n }, 
    \end{equation}
    where the summation is assumed to be norm-convergent, 
    and where each $ \omega_{ {\ms{A}}, n }, \omega_{ {\ms{B}}, n }$ is assumed to be a positive normal functional. 
    We are now ready to define the relative entanglement entropy measure introduced by~\cite{Vedral_PRL1997, Narnhofer_RepMathPhys2002}. 
    \begin{mydef}
		The relative entanglement entropy of a normal state $\omega$ on $ \sA (A) ~\bar{\otimes}~ \sA (B) $ is defined by 
		$
		E (\omega) := \inf \sbracket{ S ( \omega \| \sigma ) \mid \sigma~\textrm{separable} } 
		$, 
		where $ S (\omega \| \sigma) $ is Araki's relative entropy~\cite{Araki_RIMS1976, Araki_RIMS1977}.
    \end{mydef}  
    In the case of type~I v. Neumann algebras, 
    the normal states are in one-to-one correspondence with statistical operators $\rho_{\omega}$, via 
    $ \omega (a) = \Tr (a ~\rho_{\omega}) $. 
    In that case $ S (\omega \| \sigma) = \Tr ( \rho_{\omega} \log \rho_{\omega} - \rho_{\omega} \log \rho_{\sigma})$. 
    Our main result is:
    \begin{theorem} \label{thm: REE_Dirac}
		Let $A, B$ be open subsets with compact closure\footnote{In fact, it is sufficient to choose either $A$ or $B$ to have compact closure.} of a static Cauchy surface $\sC$ in a geodesically complete, simply connected, $D$-dimensional ($ D \in \{4,8,9,10\}~\mathrm{mod}~8 $) ultrastatic spin-spacetime 
		$(\mathscr{M} = \R \times \sC, g = \mathrm{d} t^2 -h_{ij} \mathrm{d} x^{i} \mathrm{d} x^{j} ) $  such that 
		$ d := \dist (A, B) > 0 $ and\footnote{Our sign convention for the scalar curvature $R$ is such that 
    	it is positive on the sphere.} 
		$ \inf (m^{2} + R (x) / 4) > 0 $  holds on $ (\sC, h) $,  
		where $R$ is the scalar curvature of $ (\sC, h) $ and $m$ the mass of the Dirac field.
		Then
		\begin{equation}
	    	E (\omega_{0}) \leq \mathrm{cst}. \, \exp \fbracket{ - \frac{M d}{2} }, 
	    	\quad M:= \tbracket{ \inf \fbracket{ m^{2} + \frac{1}{4} R (x)} }^{1/2}, 
	    	\quad d \to \infty,
		\end{equation}
		where the constant only depends on the geometry within an arbitrary small neighborhood of $A$ (or within $B$) and so is independent of $d$. 
    \end{theorem}
    One of our main tools in the proof will be Tomita-Takesaki theory. 
    We describe the relevant concepts and results in the next section.    
    
    \section{Tomita-Takesaki theory}
    \label{sec: Tomita_Takesaki}
    A general references to Tomita-Takesaki theory is e.g.~\citep{KadisonRingrose_II_AMS1997}. 
    To apply the theory in its most basic form we need a state 
    $\omega$ such that $\Omega_{\omega}$ is cyclic for $ \sR_{\omega} (V) $. 
    If this is the case, 
    we can define $ \TomitaVrep $  on a dense domain of $ \sF_{\omega} $ as the closure of the operator 
    \begin{equation}
		\TomitaVrep (a ~\Omega_{\omega}) = a^{*} \Omega_{\omega}, 
    \end{equation}
    where $ a \in \sR_{\omega} (V) $. 
    It is common to consider the polar decomposition $ \TomitaVrep = \JVrep \sqrt{ \DeltaVrep} $
    into an anti-unitary, involutive operator $\JVrep$ and a non-negative self-adjoint operator $\DeltaVrep$.

    If $\omega$ is quasifree, these operators are given in second quantized form~\cite{Foit_RIMS1983, Baumgartel_JMP2002}
    \begin{equation}
		\TomitaVrep = \bigoplus_{n=0}^\infty \bigwedge^n \tomitaVrep^{} , 
		\quad 
		\DeltaVrep = \bigoplus_{n=0}^\infty \bigwedge^n \deltaVrep^{} , 
		\quad
		\JVrep = \bigoplus_{n=0}^\infty \bigwedge^n \jVrep^{} , 
    \end{equation}
    densely defined on the Fermonic Fock space $ \sF_{\omega} $. 
    The 1-particle operators are conveniently described in terms of ``standard subspaces''~\cite{Longo, Foit_RIMS1983}. 
    Define 
    \begin{equation} \label{eq: def_shV}
		\sh_{\omega} (V) := \overline{ \{ \IwV ~k ~|~ k \in \sK(V), ~ \Gamma k = k \}}, 
    \end{equation}
    which is a real linear subspace of $ \sH_{\omega}$.  
    Here and in the following, we set $ \sK (V) = L^{2}(\spinorBun \! \upharpoonright V) \subset \sK $ 
    and $ \IwV := \Iw \! \upharpoonright \! \sK(V) \equiv \PV \Iw \PV $, where $\PV$ is the  projector from $\sK$ to $ \sK (V) $.  
    The subspace is called standard if $ \sh_{\omega} (V) + \ri \sh_{\omega} (V) = \sH_{\omega} $ and $ \sh_{\omega} (V) \cap \ri \sh_{\omega} (V) = \{0\}$. 
    This condition is equivalent to the condition that $ \Omega_{\omega} $ is cyclic and separating for $ \sR_{\omega} (V) $. (For the ground state $\omega_0$
    considered below, this was shown in~\cite{Strohmaier_CMP2000}.)
    Then the 1-particle Tomita operator is
    \begin{equation}
		\tomitaVrep : \sh_{\omega} (V) + \ri \sh_{\omega} (V) \to \sh_{\omega} (V) + \ri \sh_{\omega} (V), ~ 
		h' + \ri h'' \mapsto h' - \ri h''.
    \end{equation}
    The operator is closable and we denote its closure by the same symbol. 
    Furthermore $ \tomitaVrep = \jVrep \sqrt{\deltaVrep} $. 
    Another way to characterize $ \tomitaVrep $ is ($ k \in \sK (V) $)
    \begin{equation}
		\tomitaVrep (\IwV k) = \IwV (\Gamma k). 
    \end{equation}

    We will now define an unbounded operator $M_{\omega,\ms{V}}$ on $\sK (V)$ which will be subsequently related to the 
    $1$-particle modular operator. 
    The restriction of the operator $\Sigma_{\omega}$ to $ \sK (V) $ is defined by 
    $ \SigmaOmegaV := \PV \Sigma_{\omega} \PV $.  
    Then the restricted $1$-particle modular operator is characterized by:
    \begin{proposition} \label{prop: 5.2}
		Let us consider the unbounded (self-adjoint) operator 
		\begin{equation} 
	    	M_{\omega, \ms{V}} := \frac{ \PV - \SigmaOmegaV }{ \PV + \SigmaOmegaV }
		\end{equation}
		and $ f : \R_{\geq 0} \to \R $ be any continuous function. 
		Then $ f (\delta_{\omega, V}) \circ \IwV = \IwV \circ f (M_{\omega, \ms{V}})$.
    \end{proposition}
    \textbf{Proof}: 
    The proposition is proved in a different context (massive Klein-Gordon QFT) in~\cite[Prop. 5.2]{axioms5_5}. 
    The steps are rather similar, although the precise form of the operator
    $ \Sigma_{\omega} $ is different, which will concern us only later. 
    We have with $k, k' \in \sK(V)$:
    \begin{eqnarray}
		\fbracket{ \deltaVrep^{1/2} \IwV k', \deltaVrep^{1/2} \IwV k }_{\omega}
		& = & 
		\fbracket{ \tomitaVrep \IwV k, s_{\omega, V} \IwV k'}_{\omega}
		\nonumber \\ 
		& = & 
		\fbracket{ \IwV \Gamma k,  \IwV \Gamma k' }_{\omega}
		\nonumber \\ 
		& = & 
		\tfrac{1}{2} \scalarProd{ \Gamma k }{ (P_{V} + \Sigma_{\omega, V}) \Gamma k' }  
		\nonumber \\ 
		& = & 
		\tfrac{1}{2} \scalarProd{ k' }{ (P_{V} - \Sigma_{\omega, V}) k } 
		\nonumber \\ 
		& = & 
		\tfrac{1}{2} \scalarProd{ M_{\omega, \ms{V}}^{1/2} (P_{V} + \SigmaOmegaV)^{1/2} k' }{ 
	    M_{\omega, \ms{V}}^{1/2} (P_{V} + \SigmaOmegaV)^{1/2} k } 
		\nonumber \\ 
		& = & 
		\fbracket{ M_{ \omega, \ms{V} }^{1/2} U_\omega \IwV k' , M_{ \omega, \ms{V} }^{1/2} U_\omega \IwV k }_{\omega} 
		\nonumber \\ 
		& = & 
		\fbracket{U_\omega^* M_{\omega, \ms{V}}^{1/2} U_\omega \IwV k' , U_\omega^* M_{\omega, \ms{V}}^{1/2} U_\omega \IwV k }_{\omega} 
    \end{eqnarray}
    Setting $k=k'$, this gives $ \norm{ U_\omega^* M_{\omega, \ms{V}}^{1/2} U_\omega h } = \norm{ \deltaVrep^{1/2} h } $
    on the domain of $\deltaVrep$. 
    Since both operators are self-adjoint, it follows that 
    $ U_\omega^* M_{\omega, \ms{V}}^{1/2} U_\omega = \deltaVrep^{1/2} $. 
    Applying the function $f$ to this identity and $\Iw$ on the right gives the statement of the theorem. 
    \hfill \textbf{QED}
    \newline

    We will also need the following proposition:
    \begin{proposition} \label{prop: 5.3_LS}
		Let $\omega$ be any quasifree state on $ \sD (\sC)$ and define $ c_{a} := 2^{2a}$ for $ 0 < a < 1 / 4 $. 
		Then for all $ p, a \in \R_{+} $ and $ \check{V} \Subset \hat{V} \subset \sC $:
		\begin{enumerate}
	    	\item[(a)] $ \norm{\deltaVLargerep^{a} \! \upharpoonright \! \sh_{\omega} (\check{V})}_{\R, p} 
	    		= 
	    		\norm{\deltaVLargerep^{ \frac{1}{2} - a } \! \upharpoonright \! \sh_{\omega} (\check{V})}_{\R, p} $ 
	    
	    	\item[ (b) ]  $ \norm{\deltaVLargerep^{a} \! \upharpoonright \! \sh_{\omega} (\check{V})}_{\R, p} 
	    		\leq c_{a} \norm{ \fbracket{ \PVlarge - \SigmaOmegaVLarge^{2} }^{a} 
				\! \upharpoonright \! \sK (\check{V}) }_{p} $.
	\end{enumerate}
	Here the functional calculus with the convention $ 0^{a} = 0 $ has been used.
    \end{proposition}
    \begin{remark} \label{remark_lp}
		A linear map $ \Xi : \sB_{1} \to \sB_{2} $ between two Banach spaces $ \sB_{1} $ and $ \sB_{2} $ 
		is called  $p$-nuclear for $0<p\le 1$ if there exist bounded linear functionals $ \sB_{1}^{\prime} \ni \varphi_{n}, n \in \mathbb{N} $ 
		on $ \sB_{1} $ and vectors $ b_{n} \in \sB_{2} $ such that 
		\begin{equation} \label{eq: decom_p_nuclearity}
	    	\Xi (b) = \sum_{ n = 1 }^{\infty} \varphi_{n} (b)~ b_{n} \quad \forall b \in \sB_{1}, 
	    	\quad \mathrm{and} \quad 
	    	\sqrt[p]{ \sum_{ n = 1 }^{\infty} || \varphi_{n} ||^{p}~ || b_{n} ||^{p} } < \infty.
		\end{equation}
		In this case, the infimum over all decomposition~(\ref{eq: decom_p_nuclearity}) is 
		called the $p$-nuclear norm of $\Xi$:
		\begin{equation}
	    	\norm{\Xi}_{p} := \inf \sqrt[p]{ \sum_{ n = 1 }^{\infty} || \varphi_{n} ||^{p}~ || b_{n} ||^{p} }.
		\end{equation}
        We will need the following facts about this norm\footnote{It satisfies the usual axioms of a norm only for $p=1$, the case of interest for this paper. 
        For $p<1$, only a weaker version of these axioms hold.}.  
		If $ \Xi_{2} : \sB_{1} \to \sB_{2} $ and $ \Xi_{3} : \sB_{1} \to \sB_{3} $ are bounded linear maps 
		such that $ \norm{ \Xi_{3} (b) } \leq \norm{ \Xi_{2} (b) } $ for all $ b \in \sB_{1} $, 
		then $ \norm{ \Xi_{3} }_{p} \leq \norm{ \Xi_{2} }_{p} $~\cite[Lem. 2.1]{axioms5_5}. 
		One can define the real linear $\ell^{p}$ maps in an analogous way by allowing the 
		$ \varphi_{n} $ to be real linear only. 
		It follows that if a complex linear map is viewed as real linear only, then  
		$ \norm{\Xi}_{\R, p} \leq \sqrt[p]{2} \norm{\Xi}_{p} $~\cite[Remark 2.2]{axioms5_5}.
    \end{remark}
    \textbf{Proof}: 
    The proof is again rather similar to that of~\cite[Prop. 5.3]{axioms5_5}. 
    Let $ \ck (\check{V}) $ be the real subspace of $ \sK (\check{V}) $ consisting of elements $k$ such that $ \Gamma k = k $.
    Then for any $ a \in \R $ and $ k \in \ck (\check{V}) $, 
    we have, using Proposition~\ref{prop: 5.2} and 
    $ \Gamma (\PVlarge + \SigmaOmegaVLarge) \Gamma = \PVlarge - \SigmaOmegaVLarge $:
    \begin{eqnarray}\label{eqn1}
		\fbracket{ I_{ \omega, \ms{ \check{V} } } k, \deltaVLargerep^{2a} I_{ \omega, \ms{ \check{V} } } k}_{\omega}
		& = & 
		\frac{1}{2} \scalarProd{ k }{ (\PVlarge + \SigmaOmegaVLarge) M_{\omega, \ms{ \hat{V} }}^{2a}  k } 
		\nonumber \\ 
		& = & 
		\frac{1}{2} \scalarProd{ k }{ (\PVlarge + \SigmaOmegaVLarge)^{1-2a} (\PVlarge - \SigmaOmegaVLarge)^{2a}  k }
		\nonumber \\ 
		& = & 
		\frac{1}{2} \scalarProd{ k }{ (\PVlarge - \SigmaOmegaVLarge)^{1-2a} (\PVlarge + \SigmaOmegaVLarge)^{2a}  k }
		\nonumber \\ 
		& = & 
		\fbracket{ I_{ \omega, \ms{ \check{V} } } k, \deltaVLargerep^{1-2a} I_{ \omega, \ms{ \check{V} } } k }_{\omega}
    \end{eqnarray}
    The first and last terms imply 
    $ \norm{ \deltaVLargerep^{a} h }_{\omega} = \norm{ \deltaVLargerep^{1/2-a} h }_{\omega} $ 
    for all $ h \in \sh_{\omega} (\check{V}) $ so by Remark~\ref{remark_lp} item a) follows.

    Averaging over the second and third line of (\ref{eqn1}) gives 
    \begin{equation} \label{eq: norm}
	        \norm{ \deltaVLargerep^{a} I_{ \omega, \ms{ \check{V} } } k }_\omega 
		= \frac12
		\norm{ 
	    [ (\PVlarge + \SigmaOmegaVLarge)^{1 - 4a} + (\PVlarge - \SigmaOmegaVLarge)^{1 - 4a} ]^{1/2} 
	    (\PVlarge - \SigmaOmegaVLarge^{2})^{a}
	    k }_{\sK} 
    \end{equation} 
    To show item b), we note the operator inequalities 
    \begin{subequations}
		\begin{eqnarray}
	    	2^{1 - 4 a} \PVlarge \leq (\PVlarge + \SigmaOmegaVLarge)^{1 - 4 a} + (\PVlarge - \SigmaOmegaVLarge)^{1 - 4 a} 
	    	& \leq & 2 \PVlarge, \quad 0 \leq a < \frac{1}{4} 
	    	\\ 
	    	(\PVlarge + \SigmaOmegaVLarge)^{1 - 4 a} + (\PVlarge - \SigmaOmegaVLarge)^{1 - 4 a} 
	    	& = & 2, \quad a = \frac{1}{4}.  
		\end{eqnarray}
    \end{subequations}
    Using these in~\eqref{eq: norm} and applying again the Remark~\ref{remark_lp} we get
    \begin{equation}
		\norm{ \deltaVLargerep^{a} \upharpoonright \! \sh_{\omega} (\check{V}) }_{\R, p} 
		\leq 
		\mathrm{cst}. \, \norm{ (\PVlarge - \SigmaOmegaVLarge^{2})^{a} \upharpoonright \! \ck (\check{V}) }_{\R, p},
    \end{equation} 
    where on the real subspace $ \ck (\check{V}) $ of $ \sK (\check{V}) $,  the mapping $\sqrt{2} I_{ \omega, \ms{ \check{V} } }  : \ck (\check{V}) \to \sh_{\omega} (\check{V}) $
    is norm preserving (since $ \scalarProd{ k }{ \SigmaOmegaVLarge k } =  0 $), 
    and so item b) follows in view of 
    $ \norm{ (\PVlarge - \SigmaOmegaVLarge^{2})^{a} \upharpoonright \! \ck (\check{V}) }_{\R, p}
    \leq
    \norm{ (\PVlarge - \SigmaOmegaVLarge^{2})^{a} \upharpoonright \! \sK (\check{V}) }_{p} $. 
    \hfill \textbf{QED}
    
    \section{Proof of main theorem}
    Following~\cite{HollandsSanders_2017}, we proceed by first considering the modular operator 
    associated with the algebra $\sA(B')$ and the ground state $\Omega_0$ 
    (here $ B' := \sC \setminus \clo{B} $ denotes the complement of $B$ inside the Cauchy surface $\sC$). 
    It follows from the Reeh-Schlieder theorem (proved in the present context by~\cite{Strohmaier_CMP2000}) that $\Omega_0$ is cyclic\footnote{For the even part of the Fock-space, 
	to which we restrict here.} 
    and separating for $\sA(B')$. 
    We call $\DeltaBvac$, the modular operator associated with the pair ($ \sA (B'), \Omega_{0} $) and then consider the map
    \begin{equation}
		\Xi: \sA(A) \to \sF_{0} \ , \quad \Xi (a) := \DeltaBvac^{1/4} ~a ~\Omega_{0}. 
    \end{equation} 
    It has been shown that~\cite{HollandsSanders_2017} 
    \begin{equation}   \label{2}
		E (\omega_{0}) \leq \log \norm{\Xi}_{1} ,     
    \end{equation}    
    so we need to find an upper bound on the 1-norm of $\Xi$. 
    To do this, we will use the time-reversal operator, $\fT$, characterized by the following lemma.
    \begin{lemma} \label{lem: T_H_Gamma}
		In dimension $D=4,8,9,10$ mod $8$, there exists an anti-unitary operator 
		\begin{equation} \label{eq: def_T1}
	    	\fT : \sK \to \sK
		\end{equation}
		acting on (pseudo-) Majorana spinors   
		such that  $\fT^2 = +1$ if $ D = 8,9,10 ~\mathrm{mod}~8 $ and $\fT^2 = -1$ if $ D = 4 ~\mathrm{mod}~8 $, as well as
		\begin{equation} \label{eq: time_reversal_Cliff}
	    	\fT \gamma (e^{0}) \fT^{- 1} = \gamma (e^{0}), 
	    	\quad
	    	\fT \gamma (e^{i}) \fT^{- 1} = - \gamma (e^{i}), \quad i = 1, \dots, D-1.
		\end{equation} 
		$\fT$ commutes with the Dirac Hamiltonian $ H $ 
		and the conjugation $ \Gamma $.   
		Furthermore, $\fT$ preserves localization in the sense that $\fT \sK(V) = \sK(V)$ 
        and it does not depend on the choice of local frame except $e^{0}$. 
    \end{lemma}
    \begin{remark}
		It has been noted occasionally in the literature that there appears to be a problem with the  time-reversal operator $\fT$  in 
		dimension $D=3$, see~\cite{Shimizu_PTP1985, WinklerZulicke_2015}. 
		In fact, we argue below in the proof that it does not exist for any $D=3$ mod $8$.
    \end{remark}
    \textbf{Proof}: 
    Without loss of generality, we can write the desired operator $\fT = \fU \Gamma$, 
    where $\fU$ is some complex linear map on $\spinorBunCauchy$ to be defined 
    and $\Gamma$ is the Majorana conjugation, namely $\Gamma k := \bar \fC \bar k = \fC^{-1} \bar k$.  
    Since $\fT$ is to be anti-unitary, it follows that $\fU$ has to be unitary with respect to the inner product~\eqref{eq: L2_inner_product}. 
    Let $ e^{a}, a = 0, 1, \dots, D-1 $ be a chosen oriented, time-oriented orthonormal frame and 
    $ \gamma(e^a) $ the the corresponding Clifford generators. 
    It can be shown~\cite{KugoTownsend_NuclPhysB1983, Trautman_JGP2008} that 
    $\fC^{-1} \bar{\gamma} (e^{a}) \fC = \epsilon(D) \gamma (e^{a}) $ for all $a$, 
    where $ \epsilon (D) = - 1 $ if  $ D = 3, 4, 10 ~\mathrm{mod}~8 $ and $ \epsilon (D) = + 1 $ if $ D = 8,9 ~\mathrm{mod}~8 $. 
    In order for $\fT$ to have the desired commutation relation with $H$, $\fU$ must be covariantly constant on $\sC$ and 
    in order for $\fT$ to satisfy ~\eqref{eq: time_reversal_Cliff}, $\fU$ must also satisfy $ \gamma^{a} \fU \gamma^{a} = \epsilon(D) ~\fU $ for all $a$. 
    $\fU$ is now defined in a  case-by-case manner depending on the dimension $D$. 
    To state the result, we use the chirality operator 
    $ \gamma^{D+1} := \ri \gamma (e^{0}) \gamma (e^{1}) \dots \gamma (e^{D-1}) $ in all even dimensions $D$. 
    Note that the chirality operator is globally defined and independent of our choice of orthonormal frame. 
    Then we have the following cases:
    \begin{enumerate}
		\item $ D = 4 $ mod $8$: Here we must choose $ \fU = \gamma (e^{0}) \gamma^{D+1} $ and it follows that $ \fT^{2} = - 1 $. 
	
		\item $ D = 10 $ mod $8$: Here we must choose again $ \fU = \gamma (e^{0}) \gamma^{D+1}$, but now it follows that $ \fT^{2} = 1 $. 
	
		\item $ D = 8, 9 $ mod $8$: Here we must choose $ \fU = \gamma^{0} $ and it follows that $\fT^{2} = 1$.
	
		\item $D=3$ mod $8$: In these cases, an operator $\fU$ with the desired properties does not exist. 
	    	This can be seen as follows. 
	    	First, we can write any linear endomorphism as a linear combination $ \fU = \sum_{[A]} c_{[A]} \gamma (e^{[A]}) $, 
        	where $ [A] := [ a_{1} \dots a_{n} ] $ is a totally antisymmetric combination of indices 
        	and $n$ ranges from $0$ (in which case $ \gamma (e^{[A]}) = 1 $) to $ (D-1) / 2 $~\cite[Sec. 3.1.7]{FreedmanProeyen_CUP2012}. 
        	$\fU$ must also satisfy $ \gamma (e^{a}) \fU \gamma (e^{a}) = - \fU $ for all $a$. 
        	However, it can be seen by using the Clifford algebra that there is no combination of coefficients satisfying this condition except for $c_{[A]}=0$, 
        in which case $\fU$ obviously cannot be unitary. 
    \end{enumerate}
    In all cases 1.-3. it can be seen that $ [ \Gamma, \fT ] = 0 $.  
    Furthermore, $ [H, \fT] = 0 $ from~\eqref{eq: time_reversal_Cliff} and the fact that $e^{0}$ is covariantly constant on $\sC$ on an ultrastatic spacetime.   
    In all cases, $\fT$ clearly acts fibre-wise in $\spinorBunCauchy$, so $ \fT \sK (V) = \sK (V) $ follows.
    \hfill \textbf{QED}
    \newline
    
    Since $\fT^2=-1$ in $D=4$ mod $8$, while $\fT^2= 1$ in the other dimensions, these cases have to be treated somewhat differently. 
    For definiteness, we describe the argument for $D=4$ mod $8$, where a ``doubling procedure'' is necessary for our construction. 
    The doubling corresponds to going from Majorana spinors to Dirac spinors. 
    The procedure consists of going from $\sK$ to $\sK \oplus \sK$, from $\Gamma$ to $\Gamma \oplus \Gamma$, from $H$ to $H \oplus H$ etc. 
    On the doubled space, we now define
    \begin{equation} \label{eq: def_T}
		T : \sK \oplus \sK \to \sK \oplus \sK, ~k \oplus k' \mapsto T ( k \oplus k' ) := \fT k' \oplus -\fT k.
    \end{equation}
    It follows from the Lemma~\ref{lem: T_H_Gamma} that $T$ is anti-unitary, involutive ($T^2=1$), and $[T,H] = 0 = [T,\Gamma]$, 
    where $H$ means the doubled $H \oplus H$, and similarly for $\Gamma$. 
    (In the other dimensions, $T$ is defined by $ k \oplus k' \mapsto \fT k \oplus \fT k' $, 
    resulting in the same abstract properties which are all that is needed in the sequel). 
    In what follows, we will always refer to the doubled spaces/maps, so by abuse of notation, we will write 
    $\Gamma$ for $\Gamma \oplus \Gamma$, $\sK$ for $\sK \oplus \sK$ etc. 
    Physically, this means that we work with Dirac fermions. 
    Note, however, that the entanglement entropy of the doubled system is twice that of the original system, 
    because it corresponds to a (graded) tensor product on the level of the v. Neumann algebras, 
    and because our entanglement measure is sub-additive under tensor product (for details, see e.g. Section 3.2 and 3.4 of~\cite{HollandsSanders_2017}). 
    Thus, our upper bounds also apply to Majorana fermions.

    Next we prove the following lemma: 
    \begin{lemma} \label{lem: 1p_Tomita_Dirac_com_double_time_reversal}
		The $1$-particle modular operator $\deltaVvac$ (associated with the ground state $\omega_{0}) $
		commutes with $T$ and $ T \sh_{0} (V) = \sh_{0} (V) $, where $ V \subset \sC $ be an open subset.
    \end{lemma}
    \textbf{Proof}: 
    First we note that $T$ leaves $\sH_0$ invariant. 
    This follows because $\sH_0$ is the positive spectral subspace of the 1-particle Dirac Hamiltonian $H$, 
    and, as we have already proved in Lemma~\ref{lem: T_H_Gamma}, $H$ commutes with $T$. 
    The standard real subspaces $\sh_0(V)$ are invariant because $T$ commutes with the 
    Majorana conjugation operator $\Gamma$, as shown in Lemma~\ref{lem: T_H_Gamma}. 
    Thirdly, by definition for all $ h', h'' \in \sh_{0} (V) $:
    \begin{equation}
		T \tomitaVvac ( h' + \ri h'' )  
		= 
		T ( h' - \ri h'') 
		= 
		T h' + \ri T h''   
		=  
		\tomitaVvac (T h' - \ri T h'') 
		= 
		\tomitaVvac T ( h' + \ri h'' )  . 
    \end{equation}
   It follows that $ T \sh_{0} (V) \subset \sh_{0} (V) = T^2 \sh_{0} (V)\subset T \sh_{0} (V) $, because $T^2=1$, so
   $ T \sh_{0} (V) = \sh_{0} (V) $. From this it follows by construction that $T$ commutes with $\tomitaVvac$ and with $\deltaVvac$. 
   This completes the proof. 
    \hfill \textbf{QED}
    \newline

    We next define the real linear and self-adjoint projectors $ T_{\pm} := ( 1 \pm T ) / 2 $
    and the two closed complex linear subspaces 
    $ \sH_{0}^{\pm} (V) := T_{\pm} \sh_{0} (V) + \ri T_{\pm} \sh_{0} (V) \subset \sH_{0} $ 
    invariant under $T$.
    The corresponding complex orthogonal projectors are called $ E_{\ms{V}}^{\pm} : \sH_{0} \to \sH_{0}^{\pm} (V) $. 
    We also define the real orthogonal projector $ E_{\ms{V}} : \sH_{0} \to \sh_{0} (V) $.
    The latter is related to the former by $ E_{\ms{V}} = T_{+} E_{\ms{V}}^{+} + \ri  T_{-} E_{\ms{V}}^{-} $. 
    Application of Theorem 3.11 in~\cite{axioms5_5} to $ V = B' $ gives the upper bound  
    \begin{equation}
		\log \norm{\Xi}_{1}
		\leq
		2 \norm{ E_{\ms{A}}^{+} \deltaBvac^{1/4} }_{1} + 2 \norm{ E_{\ms{A}}^{-} \deltaBvac^{1/4} }_{1}. 
		\label{eq: log_3_32_LS}
    \end{equation}
    We now employ the $p$-norm inequality~\cite[Thm. 3.5]{axioms5_5}: 
    $ \norm{ E_{ \ms{A} }^{\pm} \deltaBvac }_{p} \leq {\rm cst.} \norm{ E_{ \ms{A} } \deltaBvac }_{\R, p} $, 
    where the numerical constant cst. depends only on $p$, 
    and then our Proposition~\ref{prop: 5.3_LS} gives ($ \hat{V} = B', \check{V} = A $):
    \begin{eqnarray}
		\log \norm{\Xi}_{1}
		& \leq & 
		\mathrm{cst}. \, \norm{ \deltaBvac^{1/4} \! \upharpoonright \! \sh_{0} (A) }_{\R, 1} 
		\nonumber \\ 
		& \leq & 
		\mathrm{cst}. \, \norm{ (P_{ \ms{B}' } - \Sigma_{ 0, \ms{B}' }^{2} )^{1/4} \! \upharpoonright \! \sK (A) }_{1}.
    \end{eqnarray}

    We summarize our findings so far in the following lemma:
    \begin{lemma} \label{lem: REE_1_norm}
		We have 
		\begin{equation}
	    	E (\omega_0) 
	    	\leq 
	    	\mathrm{cst}. \, \norm{ (P_{ \ms{B}' } - \Sigma_{ 0, \ms{B}' }^{2} )^{1/4} \! \upharpoonright \! \sK (A) }_{1}
		\end{equation}
		where $ \mathrm{cst}. $ is a numerical constant independent of $A$ and $B$. 
    \end{lemma}
    Our proof of the main theorem thus boils down to a norm estimation of the operator 
    $ \left(P_{ \ms{B}' } - \Sigma_{ 0, \ms{B}' }^{2} \right)^\frac14 \! \upharpoonright \! \sK (A) $, 
    which occupies the remainder of this section. 
    We introduce two intermediate regions between $A$ and $B$, called $\check{V}$ and $\hat{V}$. 
    These regions are such that $ A \subset \check{V} \subset \hat{V} \subset B' $. We then introduce
    smooth functions $ \check{\chi}, \hat{\chi}, \chi $ such that: 
    1) $\check{\chi}$ is supported in $\check{V}$ and $\check{\chi} \equiv 1$ on $A$. 
    2) $\hat{\chi}$ is supported in $B'$ and $ \hat{\chi} \equiv 1 $ on $\hat V$, $ 1- \hat{\chi} \equiv 1 $ on $B$. 
    3) The distance $ \dist (\supp{\check{\chi}}, \supp{1-\hat{\chi}} ) = d - \varepsilon $, where
    $ d =\dist (A,B) $ and $ \varepsilon > 0 $ is thought of as small.  
    4) $\chi$ is a function of compact support such that $ \chi \equiv 1$ on $\supp{\check{\chi}}$, 
    $\supp{\chi} \subset \supp{\hat{\chi}}$ and $ \dist( \supp{\chi}, \supp{1- \hat{\chi}} ) = d - 2 \varepsilon $. 
    The situation is sketched in the following Fig.~\ref{fig: support}. 
    
    \begin{center}
	\begin{tikzpicture}
	    \draw[ultra thick, >={Triangle[length=2mm,width=2mm]}, ->] (-2.5, 0) -- (-3, 0);
	    \draw[line width=1pt, >={Triangle[length=2mm,width=2mm]}, ->] (0.5, 0) -- (0.5, 1.75);
	    \draw[green, ultra thick] (-2.5, 0) -- (0.5, 0);
	    \draw[red, ultra thick] (7, 0) -- (9, 0);
	    \draw[blue, ultra thick] (6.5, 0) -- (7, 0);
	    \draw[blue, ultra thick] (9, 0) -- (9.5, 0);
	    \draw[brown, ultra thick] (2, 0) -- (6.5, 0);
	    \draw[brown, ultra thick, >={Triangle[length=2mm,width=2mm]}, ->] (9.5, 0) -- (12, 0);
	    \draw[black, ultra thick] (0.5, 0) -- (2, 0);
	    \draw[blue, thick] (7, 1) -- (9, 1);
	    \draw[blue, thick] (7, 1) to [out=190, in=20] (6.5, 0);
	    \draw[blue, thick] (9, 1) to [out=350, in=160] (9.5, 0);
	    \draw[brown, thick] (6.5, 1) -- (2, 1);
	    \draw[brown, thick] (9.5, 1) -- (12, 1);
	    \draw[brown, thick] (2, 1) to [out=185, in=20] (0.5, 0);
	    \draw[brown, thick] (-3, 0.5) to [out=355, in=150] (-2.5, 0);
	    \draw[cyan, thick] (6.5, 1) -- (7, 1);
	    \draw[cyan, thick] (9.5, 1) -- (9, 1);
	    \draw[cyan, thick] (6.5, 1) to [out=190, in=20] (6, 0);
	    \draw[cyan, thick] (9.5, 1) to [out=350, in=160] (10, 0);
	    \draw[green, thick] (-2.5, 1) -- (0.5, 1);
	    \draw[green, thick] (-2.5, 1) to [out=190, in=20] (-3, 0.5);
	    \draw[green, thick] (0.5, 1) to [out=355, in=160] (2, 0);
	    \node[below] at (8, 0) {$ \textcolor{red}{ A } $};
	    \node[above] at (8, 1) {$ \textcolor{blue}{ \check{\chi} } $};
	    \node[below] at (6.75, 0) {$ \textcolor{blue}{  \check{V} } $};
	    \node[above] at (6.75, 1) {$ \textcolor{cyan}{\chi} $};
	    \node[below] at (4.5, 0) {$ \textcolor{brown}{  \hat{V} } $};
	    \node[above] at (4.5, 1) {$ \textcolor{brown}{ \hat{\chi} } $};
	    \node[below] at (-1, 0) {$ \textcolor{green}{B} $};
	    \node[above] at (-1, 1) {$ \textcolor{green}{ 1 - \hat{\chi} } $};
	    \node[below] at (1.25, 0) {$ B' $};
	    \node[below] at (12, 0) {$ \sC $};
	    \node[left] at (0.5, 1.2) {$ 1 $};
	\end{tikzpicture}
	\captionof{figure}{The supports of the cutoff functions.}
	\label{fig: support}
    \end{center}
    
    We next show
    \begin{proposition} \label{prop: ground_state_lp}
		On a complete Riemannian manifold $ (\sC, h) $ we have
		\begin{equation}
	    	\norm{ (P_{ \ms{B}' } - \Sigma_{ 0, \ms{B}' }^{2} )^{1/4} \PA }_{1} 
	    	\leq 
	    	\mathrm{cst}. \, \exp\fbracket{ - \frac{M d}{2} }, 
	    	\quad d \to \infty, 
		\end{equation} 
		where $ M := [ \inf (m^{2} + R (x) / 4) ]^{1/2} $ with Dirac mass $m$ and Ricci scalar $R$.
    \end{proposition} 
    \textbf{Proof}: 
    Let $\check X$ be the multiplication operator by $\check \chi$, $\hat X$ that associated with $\hat \chi$, 
    and $X$ that associated with $\chi$. 
    We note (with $\Sigma_0 =|H|^{-1} H$, $H$ the Dirac Hamiltonian):
    \begin{eqnarray} \label{eq:36}
		\PA (P_{ \ms{B}' } - \SigmaB^2) \PA 
		& = & 
		\PA \Sigma_{0} (1-P_{ \ms{B}' }) \Sigma_{0} \PA 
		\nonumber \\
		& = & 
		| \PB \Sigma_{0} \PA |^2
		\nonumber \\
		& = & 
		| \PB (1-\hat{X}) \Sigma_{0} \check{X} \PA |^2.
    \end{eqnarray}
    Here we used the definition of $ \SigmaB = P_{ \ms{B}' } \Sigma_{0} P_{ \ms{B}' } $ 
    and the definitions of the cutoff functions. 
    We have
    \begin{equation}
		[ \PA (P_{ \ms{B}' } - \SigmaB^2)^{1/2} \PA ]^2 \leq  \PA (P_{ \ms{B}' } - \SigmaB^2) \PA
    \end{equation}
    and taking the square root and using the operator monotone property of the square root, we get 
    \begin{equation}
		| (P_{ \ms{B}' } - \SigmaB^{2})^{1/4} \PA | 
		\leq 
		[ \PA (P_{ \ms{B}' } - \SigmaB^{2}) \PA ]^{1/2} 
		= 
		| \PB (1 - \hat{X}) \Sigma_{0} \check{X} \PA |, 
    \end{equation}
    and consequently using the properties of the 1-norm and $ \norm{\PA} = \norm{ P_{ \ms{B}' } } = 1 $, and using 
    \eqref{eq:36}
    \begin{equation}\label{11}
		\norm{ (P_{ \ms{B}' } - \SigmaB^{2})^{1/4} P_A }_{1} 
		\leq 
		\norm{ P_B  (1-\hat X) \Sigma_0 \check X P_A }_{1} 
		= 
		\norm{ (1-\hat X) \Sigma_0 \check X }_{1}. 
    \end{equation} 
    Employing the Schr\"{o}dinger~\cite{Schrodinger_1932}-Lichnerowicz~\cite{Lichnerowicz_1963} formula for the square of the Dirac operator, we have $ L = H^{*} H $, 
    where $L$ is the modified Lichnerowicz Laplacian, defined by 
    \begin{equation} \label{eq: Laplace_Beltrami}
		L := - \sum_i \nabla_{ e_{i} } \nabla_{ e_{i} } + m^{2} + \frac{1}{4} R , 
    \end{equation}  
    with $ \sum \nabla_{ e_{i} } \nabla_{ e_{i} } $ the usual Lichnerowicz Laplacian on the complete Riemannian manifold $ (\sC, h) $. 
    Then using the support property of $\chi$ and the fact that $H$ is a partial differential operator
    \begin{equation}
		(1 - \hat{X}) \Sigma_{0} \check{X} 
		= 
		( 1 - \hat{X}) |H|^{-1} H \check{X} 
		= 
		( (1-\hat{X}) L^{-1/2} X L^{a} ) (L^{-a} H \check{X}). 
    \end{equation}
    Therefore, using the properties of the 1-norm
    \begin{equation}
		\norm{ (1-\hat{X}) \Sigma_{0} \check{X} }_{1} 
		\leq 
		\norm{ (1 - \hat{\chi}) L^{-1/2} \chi L^{a} }~ \norm{ L^{-a} H \check{\chi} }_{1} 
		\leq  \norm{ (1-\hat{\chi}) L^{-1/2} \chi L^{a} }~ \norm{ L^{-a+1/2} \check{\chi} }_{1}. 
    \end{equation}
    It follows from the same argument as given in Appendix A of~\cite{AntoniHollands_CMP2006} 
    that $ \norm{ L^{-a+1/2} \check{\chi} }_{1} < \infty $ provided 
    $ 2a - 1 \geq D-1 $, or $ a \geq D/ 2 $. 
    Using $ \norm{ (1-\hat{\chi}) L^{-1/2} \chi L^{a} } \leq \mathrm{cst}. \, \exp( -Md / 2) $ 
    (see Theorem~\ref{thm: 4_5_LS} in the Appendix) together with eq.~\eqref{11} completes the proof of the proposition. 
    \hfill \textbf{QED}

    In view of the Lemma~\ref{lem: REE_1_norm}, the proposition proves the main theorem. 
    \hfill \textbf{QED} 
    
    \section*{Acknowledgement}
    This work grew out of the master thesis by one of the authors (O.I.). 
    He is grateful to Christian B\"{a}r for discussions  
    during the summer school on ``Geometric Cauchy problems on Lorentzian manifolds,'' 
    at Regensburg. He acknowledges financial support by this conference. 
    
    \begin{appendix}
	\section{A theorem for the (modified) Lichnerowicz Laplacian on a complete Riemannian manifold}
	\label{sec: Laplace_Beltrami}
	\begin{theorem} \label{thm: 4_5_LS}
	    Let $ \chi \in C_{ \textrm{c} }^{\infty} (V, \R), \hat{\chi} \in C_{ \textrm{c} }^{\infty} (\sC, \R) $ and  
	    $ V, \hat{V} $ are open subsets of $\sC$ with compact closures such that $ \clo{V} \subset \hat{V} $ with  
        $ \hat{\chi} \equiv 1 $ on $ \hat{V} $. 
	    Assume that $\inf (m^{2} + R (x) / 4) > 0 $ holds on the complete Riemannian manifold $ (\sC, h) $. 
	    Then the operator $ ( 1 - \hat{\chi}) L^{a} \chi L^{b} $ is bounded for all $ a, b \in \R $ by 
	    \begin{equation}
			\norm{ ( 1 - \hat{\chi}) L^{a} \chi L^{b} }^{2} 
			\leq 
			\sum_{ n = 0 }^{N} \mathrm{cst}_{n + a}.~ M^{ -4 ( N - b ) - 3}  d^{ - 2 (n + a + 1)} \mathrm{e}^{-M d}, 
			\quad b \leq N \in \{0\} \cup \N, \, d \to \infty,  
	    \end{equation}
	    where the constants $ \mathrm{cst}_{n+a} $ are independent of  
	    $ d := \mathrm{dist} \fbracket{ \clo{ \hat{V}' }, \clo{V} }, 
	    \, \hat{V}'  := \sC \setminus \clo{ \hat{V} } $ 
	    but depend only on the geometry inside $ V, \hat{V}' $, the test-functions $\chi, \hat \chi$ and 
	    the exponents $a,b$ of the modified Lichnerowicz Laplacian~\eqref{eq: Laplace_Beltrami} $L$.  
	    Here $ M := [ \inf (m^{2} + R (x) / 4) ]^{1/2} $ 
	    with $m$ the Dirac mass and $R$ is the Ricci scalar of $ (\sC, h) $, 
        and $N$ the smallest natural number not smaller than $b$.
	\end{theorem}
	\textbf{Proof}: 
	We follow the strategy of Theorem 4.5 in~\cite{axioms5_5} to get our bound.
	By construction, $ V, \hat{V} \subset \sC $ are open subsets such that the supports of 
	$ 1 - \hat{\chi} $ and $ \chi $ are contained in $ \hat{V}' $ and $ V $, resp., 
	and $ \clo{ \hat{V}' } $ and $ \clo{V} $ are disjoint. 
	Furthermore, $V$ is relatively compact, so that $ \clo{ \hat{V}' } $ 
	and $ \clo{V} $ are separated by a minimal distance $ d > 0 $. 
	
	\begin{center}
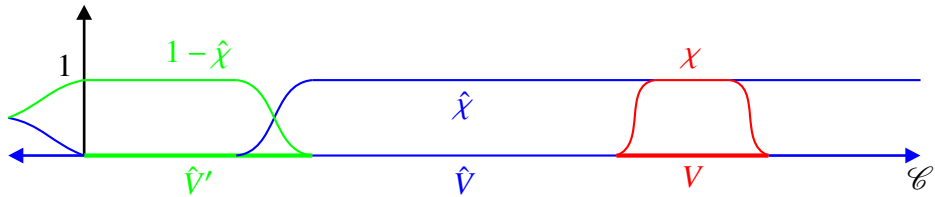

	    \begin{tikzpicture}
		\draw[blue, line width=1pt, >={Triangle[length=2mm,width=2mm]}, ->] (9, 0) -- (11, 0);
		\draw[line width=1pt, >={Triangle[length=2mm,width=2mm]}, ->] (0, 0) -- (0, 2);
		\draw[green, ultra thick] (0, 0) -- (3, 0);
		\draw[red, ultra thick] (7, 0) -- (9, 0);
		\draw[blue, thick] (3, 0) -- (7, 0);
		\draw[blue, line width=1pt, >={Triangle[length=2mm,width=2mm]}, ->] (0, 0) -- (-1, 0);
		\draw[red, thick] (7.5, 1) -- (8.5, 1);
		\draw[blue, thick] (3, 1) -- (7.5, 1);
		\draw[blue, thick] (8.5, 1) -- (11, 1);
   		\draw[blue, thick] (3, 1) to [out=185, in=5] (2, 0);
		\draw[blue, thick] (-1, 0.5) to [out=350, in=160] (0, 0);
		\draw[green, thick] (0, 1) -- (2, 1);
		\draw[green, thick] (2, 1) to [out=355, in=175] (3, 0);
		\draw[green, thick] (0, 1) to [out=190, in=20] (-1, 0.5);
		\draw[red, thick] (8.5, 1) to [out=350, in=160] (9, 0);
		\draw[red, thick] (7.5, 1) to [out=190, in=20] (7, 0);
		\node[below] at (8, 0) {$ \textcolor{red}{V} $};
		\node[above] at (8, 1) {$ \textcolor{red}{\chi} $};
		\node[below] at (5, 0) {$ \textcolor{blue}{  \hat{V} } $};
		\node[below] at (5, 1) {$ \textcolor{blue}{ \hat{\chi} } $};
		\node[below] at (1.5, 0) {$ \textcolor{green}{ \hat{V}' } $};
		\node[above] at (1.5, 1) {$ \textcolor{green}{ 1 - \hat{\chi} } $};
		\node[below] at (11, 0) {$ \sC $};
		\node[left] at (0, 1.2) {$ 1 $};
	    \end{tikzpicture}
	    \captionof{figure}{The supports of the cutoff functions.}
	    \label{Fig: support2}
	\end{center}
	
	For any $ r \in \sbracket{0} \cup \N $, let 
    $ P = \sum_{ n = 0 }^{r} \alpha_{n}^{ b_{1} \hdots b_{n} } \nabla_{ b_{1} } \cdots \nabla_{ b_{n} } $ 
    be a partial differential operator of order $r$ in the bundle $\spinorBunCauchy$ 
    whose coefficients $\alpha_{n}$ are real endomorphisms of compact support in $V$ in each fibre of the bundle
    $\spinorBunCauchy$, with respect to the fibre wise inner product provided by the integrand of~\eqref{eq: L2_inner_product}. 
    If $P$ is additionally a hermitian operator on $\LtwoCauchy$ with respect to~\eqref{eq: L2_inner_product}, 
    then one can show by induction that $r$ must be even and that $P$ can be written as 
    $ 
    P = 
    \sum_{ n = 0 }^{r/2} \nabla_{ a_{1} } \cdots \nabla_{ a_{n} } \, \xi_{n}^{ a_{1} \hdots a_{n} \, b_{1} \hdots b_{n} } \,
    \nabla_{ b_{1} } \cdots \nabla_{ b_{n} } 
    $ 
    for compactly supported (in $V$) real endomorphisms $ \xi_{n}^{ a_{1} \hdots a_{n} \, b_{1} \hdots b_{n} } $ of the bundle $\spinorBunCauchy$. 
    Applying this fact to the operator $ P = \chi L^{2N} \chi $, we get
    \begin{equation} 
	    \norm{ L^{N} \chi k }^{2} 
	    \leq 
        \sum_{n=0}^{2N}  \scalarProd{ \nabla_{ a_{1} } \cdots \nabla_{ a_{n} }  k }{ 
        \xi_{n}^{ a_{1} \hdots a_{n} \, b_{1} \hdots b_{n}} \, \nabla_{ b_{1} } \cdots \nabla_{ b_{n} } k } , 
        \quad N \in \N, 
    \end{equation}
    where the norm is that in the space $\LtwoCauchy$.
    Viewing $\xi_{n}$ as a linear map on the bundle $ \spinorBunCauchy \otimes (T \sC)^{\otimes n} $ with fibre wise norm $\norm{ \xi_{n} }$, 
    we can find a smooth real valued function $\chi_{n}$ of compact support on $V$ such that $ \| \xi_{n} \| \leq \chi_{n}^{2} $. 
    So we get
    \begin{equation} 
	    \norm{ L^{N} \chi k }^{2} 
	    \leq 
        \sum_{n=0}^{2N} \scalarProd{\nabla_{ a_{1} } \cdots \nabla_{a_n} k}{ \chi_{n}^{2} \, \nabla^{ a_{1} } \cdots \nabla^{ a_{n} } k}.  
    \end{equation}
    Consider now for even $n$ the operator 
    $ 
    P_{n} = 
    (-1)^{n} \nabla_{ a_{1} } \cdots \nabla_{ a_{n} } \, \chi_{n}^{2} \, \nabla^{ a_{1} } \cdots \nabla^{ a_{n} } 
    - L^{n/2} \chi_{n}^{2} L^{n/2}
    $ 
    and for odd $n$ the operator 
    $ 
    P_{n} = 
    (-1)^{n} \nabla_{ a_{1} } \cdots \nabla_{ a_{n} } \, \chi_{n}^{2} \, \nabla^{ a_{1} } \cdots \nabla^{ a_{n} } 
    + L^{(n-1)/2} \nabla_{a} \, \chi_{n}^{2} \, \nabla^{a} L^{(n-1)/2}
    $, 
    which have order $2n-1$, are hermitian, and have real coefficients. 
    We can therefore write 
    \begin{equation*} 
	    \norm{ L^{N} \chi k }^{2} 
	    \leq 
	    \sum_{ n = 0, \rm even }^{2N} \norm{ \chi_{n} L^{n / 2} k }^{2} 
        +
        \sum_{ n = 1, \rm odd }^{2N} \norm{ \chi_{n}  \nabla_{a} L^{(n - 1) / 2} k }^{2}
        +
        \sum_{ n = 0}^{N} \scalarProd{k}{ P_n k} . 
	\end{equation*}
    For the middle term, we can use the estimate 
    $ - \nabla_{a} \, \chi^{2}_{n} \, \nabla^{a} \leq L \, \chi_{n}^{2}  (2 f)^{-1} L + (\nabla_{a} \nabla^{a} \, \chi_{n}^{2} ) / 2 $ where $ f := m^{2} + R / 4 > 0 $.
    We can therefore write 
    $
	    \norm{ L^{N} \chi k }^{2} 
	    \leq 
	    \sum_{ n = 0}^{N} \norm{ \eta_{n} L^{n} k }^{2} +
        \sum_{ n = 0}^{N} \scalarProd{ k}{ Q_{n} k }
	$
    for some real valued smooth functions $\eta_n$ supported in $V$ and hermitian 
    partial differential operators $Q_{n}$ with real coefficients supported in $V$ of order at most $2n-1$.

    Repeating the above argument and proceeding inductively in this way, we 
    arrive at the conclusion that there are new
	$ \eta_{1}, \eta_{2}, \hdots, \eta_{N} \in C_{ \textrm{c} }^{\infty} (V, \R) $ 
	such that 
	$
	    \norm{ L^{N} \chi k }^{2} 
	    \leq 
	    \sum_{ n = 0 }^{N} \norm{ \eta_{n} L^{n} k}^{2}
	$.
    The operator $L$ is known to be essentially self-adjoint~\cite{Chernoff_JFA1973, Wolf_Indiana1973} on smooth sections of compact support.
	For any real number $a$ and compactly supported section $k$ on $\spinorBunCauchy$, 
	the vector $ L^{a} k $ is therefore in the domain of all powers of $L$, hence smooth. 
	Therefore, for $\theta$ real-valued and smooth, 
	\begin{equation} \label{eq: match5}
	    \norm{ L^{N} \chi L^{a} \theta k }^{2} 
	    \leq 
	    \sum_{ n = 0 }^{N} \norm{ \eta_{n} L^{n + a} \theta k }^{2} 
	\end{equation}
    as long as $k$ has compact support. 
	Let us now consider a $\theta$ supported in $ \hat{V}' $. 
    By definition of $M^2$, we can write $ L = Y + M^{2} $, where 
    $ Y := - \sum \nabla_{ e_{i} } \nabla_{ e_{i} } + R / 4 + m^{2} - M^{2} $ is a positive, self-adjoint operator. 
    The Fourier transform $ \tilde{F} (s) $ of the smooth function $ F (\lambda) := ( \lambda^{2} + M^{2} )^{n+a} $ 
    satisfies a bound of the form 
    \begin{equation} \label{eq: bound}
    	\int_{d}^{\infty} | \hat{F} (s) | ~\mathrm{d} s 
        \leq \mathrm{cst}_{n + a}. \, M^{-3 / 2} d^{ - (n + a + 1) } \mathrm{e}^{- M d}, 
        \quad d \to \infty. 
    \end{equation} 
    We then write
    \begin{eqnarray}
    	\norm{ \eta_{n} L^{n + a} \theta k} 
    	& = &
    	\norm{
    	\frac{1}{2\pi} \int_{-\infty}^\infty \mathrm{d} s ~\tilde{F} (s) \eta_{n} \cos( s \sqrt{Y} ) \theta k 
    	}
    	\nonumber \\ 
    	& = &
    	\norm{
    	\frac{1}{\pi} \int_{d}^{\infty} \mathrm{d} s ~\tilde{F} (s) \eta_{n} \cos( s \sqrt{Y} ) \theta k 
    	} 
    	\nonumber \\ 
    	& \leq & 
    	\frac{1}{\pi} \int_{d}^{\infty} \mathrm{d} s ~| 
        \tilde{F} (s) | \norm{ \eta_{n} }_{\infty} \norm{\theta}_{\infty} \norm{k}.
    \end{eqnarray}
    Here we have used in the second line the trick of using the finite propagation speed of the spinor wave operator
    $\partial_s^2 + Y$ following an idea of~\cite[Prop. 1.1]{Cheeger_JDG1982}. 
    Using now~\eqref{eq: bound}, we arrive at the 
	following bound generalizing~\cite[Prop. 4.3]{axioms5_5} to our setting: 
	\begin{equation} \label{eq: match6}
	    \norm{ \eta_{n} L^{n + a} \theta } 
	    \leq 
	    \mathrm{cst}_{n+a}. \, \norm{\theta}_{\infty} \norm{ \eta_{n} }_{\infty}  M^{-3 / 2} d^{ - (n + a + 1) } \mathrm{e}^{- M d}, 
	    \quad d \to \infty, 
	\end{equation}
	where the constant depends only on the exponent of $L$.  
	By the definition of the operator norm and using~\eqref{eq: match6}: 
	\begin{equation} \label{eq: match7}
	    \norm{ \eta_{n} L^{n +a} \theta k }^{2} 
	    \leq 
        \norm{\theta}_{\infty}^{2} \norm{ \eta_{n} }_{\infty}^{2} \rN (n +a)^{2} \norm{k}^{2},
	\end{equation}
	using the shorthand $ \rN(a) = \mathrm{cst}_{a}. \, M^{-3 / 2} d^{ - (a + 1)} \mathrm{e}^{- M d } $. 
	Substituting~\eqref{eq: match7} into~\eqref{eq: match5}, we obtain 
	$ 
    \norm{ \theta L^{a} \chi  L^{N} } 
	\leq 
	\norm{\theta}_{\infty}^{2} \sum_{ n = 0 }^{N} \rN (n + a)^{2} \norm{ \eta_{n} }_{\infty}^{2} 
    $.
	Finally for any real number $b$ such that $ b \leq N $ we arrive at 
	\begin{eqnarray}
	    \norm{ \theta L^{a} \chi L^{b} }^{2} 
	    =
	    \norm{ \theta L^{a} \chi L^{N} L^{ - (N - b)} }^{2} 
	    \leq 
	    \norm{\theta}_{\infty}^{2} \sum_{ n = 0 }^{N} \rN (n + a)^{2} \norm{ \eta_{n} }_{\infty}^{2} 
	    M^{ - 4 ( N - b) }. 
	\end{eqnarray}
	Absorbing the product of supremum norms of the bounded function $\theta=1-\hat \chi$ and the compactly supported functions 
	$ \sbracket{ \eta_{n} }_{n \in \N} $ into the constants $ \sbracket{ \mathrm{cst}_{n} }_{n \in \N} $  
	completes the proof. 
	\hfill \textbf{QED} 
    \end{appendix}


\begin{thebibliography}{34}%
\makeatletter
\providecommand \@ifxundefined [1]{%
 \@ifx{#1\undefined}
}%
\providecommand \@ifnum [1]{%
 \ifnum #1\expandafter \@firstoftwo
 \else \expandafter \@secondoftwo
 \fi
}%
\providecommand \@ifx [1]{%
 \ifx #1\expandafter \@firstoftwo
 \else \expandafter \@secondoftwo
 \fi
}%
\providecommand \natexlab [1]{#1}%
\providecommand \enquote  [1]{``#1''}%
\providecommand \bibnamefont  [1]{#1}%
\providecommand \bibfnamefont [1]{#1}%
\providecommand \citenamefont [1]{#1}%
\providecommand \href@noop [0]{\@secondoftwo}%
\providecommand \href [0]{\begingroup \@sanitize@url \@href}%
\providecommand \@href[1]{\@@startlink{#1}\@@href}%
\providecommand \@@href[1]{\endgroup#1\@@endlink}%
\providecommand \@sanitize@url [0]{\catcode `\\12\catcode `\$12\catcode
  `\&12\catcode `\#12\catcode `\^12\catcode `\_12\catcode `\%12\relax}%
\providecommand \@@startlink[1]{}%
\providecommand \@@endlink[0]{}%
\providecommand \url  [0]{\begingroup\@sanitize@url \@url }%
\providecommand \@url [1]{\endgroup\@href {#1}{\urlprefix }}%
\providecommand \urlprefix  [0]{URL }%
\providecommand \Eprint [0]{\href }%
\providecommand \doibase [0]{http://dx.doi.org/}%
\providecommand \selectlanguage [0]{\@gobble}%
\providecommand \bibinfo  [0]{\@secondoftwo}%
\providecommand \bibfield  [0]{\@secondoftwo}%
\providecommand \translation [1]{[#1]}%
\providecommand \BibitemOpen [0]{}%
\providecommand \bibitemStop [0]{}%
\providecommand \bibitemNoStop [0]{.\EOS\space}%
\providecommand \EOS [0]{\spacefactor3000\relax}%
\providecommand \BibitemShut  [1]{\csname bibitem#1\endcsname}%
\let\auto@bib@innerbib\@empty
\bibitem [{\citenamefont {Clifton}\ and\ \citenamefont
  {Halvorson}(2001)}]{CliftonHalvorson_HPMP2001}%
  \BibitemOpen
  \bibfield  {author} {\bibinfo {author} {\bibfnamefont {R.}~\bibnamefont
  {Clifton}}\ and\ \bibinfo {author} {\bibfnamefont {H.}~\bibnamefont
  {Halvorson}},\ }{\enquote {\bibinfo {title} {Entanglement and open systems in algebraic quantum field theory},}}\ 
  \href {\doibase
  http://dx.doi.org/10.1016/S1355-2198(00)00033-2} {\bibfield  {journal}
  {\bibinfo  {journal} {Stud. Hist. Philos. Mod. Phys.}\ }\textbf {\bibinfo
  {volume} {32}},\ \bibinfo {pages} {1 } (\bibinfo {year} {2001})},\ \Eprint
  {http://arxiv.org/abs/0001107v1} {arXiv:0001107v1 [quant-ph]}\BibitemShut
  {NoStop}%
\bibitem [{\citenamefont {Vedral}\ \emph {et~al.}(1997)\citenamefont {Vedral},
  \citenamefont {Plenio}, \citenamefont {Rippin},\ and\ \citenamefont
  {Knight}}]{Vedral_PRL1997}%
  \BibitemOpen
  \bibfield  {author} {\bibinfo {author} {\bibfnamefont {V.}~\bibnamefont
  {Vedral}}, \bibinfo {author} {\bibfnamefont {M.~B.}\ \bibnamefont {Plenio}},
  \bibinfo {author} {\bibfnamefont {M.~A.}\ \bibnamefont {Rippin}}, \ and\
  \bibinfo {author} {\bibfnamefont {P.~L.}\ \bibnamefont {Knight}},\ }{\enquote {\bibinfo {title} {Quantifying entanglement},}}\
  \href
  {\doibase 10.1103/PhysRevLett.78.2275} {\bibfield  {journal} {\bibinfo
  {journal} {Phys. Rev. Lett.}\ }\textbf {\bibinfo {volume} {78}},\ \bibinfo
  {pages} {2275} (\bibinfo {year} {1997})},\ \Eprint
  {http://arxiv.org/abs/9702027v1} {arXiv:9702027v1 [quant-ph]}\BibitemShut
  {NoStop}%
\bibitem [{\citenamefont {Hollands}\ and\ \citenamefont
  {Sanders}(2017)}]{HollandsSanders_2017}%
  \BibitemOpen
  \bibfield  {author} {\bibinfo {author} {\bibfnamefont {S.}~\bibnamefont
  {Hollands}}\ and\ \bibinfo {author} {\bibfnamefont {K.}~\bibnamefont
  {Sanders}},\ }{\enquote {\bibinfo {title} {Entanglement measures and their properties in quantum field theory},}}\
  \href {https://arxiv.org/abs/1702.04924v2} {\bibfield
  {journal} {\bibinfo  {journal} {arXiv: 1702.04924v2 [quant-ph]}\ } (\bibinfo
  {year} {2017})}\BibitemShut 
  {NoStop}%
\bibitem [{\citenamefont {Donald}\ \emph {et~al.}(2002)\citenamefont
  {Donald}, \citenamefont {Horodecki},\ and\ \citenamefont
  {Rudolph}}]{Donald_PLA1999}%
  \BibitemOpen
  \bibfield  {author} {\bibinfo {author} {\bibfnamefont {M. J.}~\bibnamefont
  {Donald}}, \bibinfo {author} {\bibfnamefont {M.}~\bibnamefont {Horodecki}},
  \ and\ \bibinfo {author} {\bibfnamefont {O.}~\bibnamefont {Rudolph}},\ }{\enquote {\bibinfo {title} {The Uniqueness Theorem for Entanglement Measures},}}\
  \href
  {\doibase 10.1063/1.1495917} {\bibfield  {journal} {\bibinfo  {journal} {J.
  Math. Phys.}\ }\textbf {\bibinfo {volume} {43}},\ \bibinfo {pages} {4252}
  (\bibinfo {year} {2002})},\ \Eprint {https://arxiv.org/abs/quant-ph/0105017v2}
  {arXiv: 0105017v2 [quant-ph/]} \BibitemShut 
  {NoStop}%
\bibitem [{\citenamefont {Calabrese}\ \emph {et~al.}(2009)\citenamefont
  {Calabrese}, \citenamefont {Cardy},\ and\ \citenamefont
  {Tonni}}]{Calabrese_JStatMech2009}%
  \BibitemOpen
  \bibfield  {author} {\bibinfo {author} {\bibfnamefont {P.}~\bibnamefont
  {Calabrese}}, \bibinfo {author} {\bibfnamefont {J.}~\bibnamefont {Cardy}},
  \ and\ \bibinfo {author} {\bibfnamefont {E.}~\bibnamefont {Tonni}},\ }{\enquote {\bibinfo {title} {Entanglement entropy of two disjoint intervals in conformal field theory},}}\
  \href
  {http://iopscience.iop.org/article/10.1088/1742-5468/2009/11/P11001/meta#citations} {\bibfield  {journal} {\bibinfo  {journal} {J.
  Stat. Mech.}\ }\textbf {\bibinfo {volume} {0911}},\ \bibinfo {pages} {11001}
  (\bibinfo {year} {2009})},\ \Eprint {https://arxiv.org/abs/0905.2069v2}
  {arXiv: 0905.2069v2 [hep-th]} \BibitemShut 
  {NoStop}%
\bibitem [{\citenamefont {Casini}\ and\ \citenamefont
  {Huerta}(2009)}]{CasiniHuerta_JPA2009}%
  \BibitemOpen
  \bibfield  {author} {\bibinfo {author} {\bibfnamefont {H.}~\bibnamefont
  {Casini}}\ and\ \bibinfo {author} {\bibfnamefont {M.}~\bibnamefont
  {Huerta}},\ }{\enquote {\bibinfo {title} {Entanglement entropy in free quantum field theory},}}\ 
  \href {http://iopscience.iop.org/article/10.1088/1751-8113/42/50/504007/meta;jsessionid=173F63D1AC3908C51DDD5319EF0C07E9.c4.iopscience.cld.iop.org} {\bibfield  {journal}
  {\bibinfo  {journal} {J. Phys. A}\ }\textbf {\bibinfo
  {volume} {42}},\ \bibinfo {pages} {504007 } (\bibinfo {year} {2009})},\ \Eprint
  {https://arxiv.org/abs/0905.2562v3} {arXiv: 0905.2562v3 [hep-th]}\BibitemShut
  {NoStop}%
\bibitem [{\citenamefont {Rangamani}\ and\ \citenamefont
  {Takayanagi}(2017)}]{RangamaniTakayanagi_Springer2017}%
  \BibitemOpen
  \bibfield  {author} {\bibinfo {author} {\bibfnamefont {M.}~\bibnamefont
  {Rangamani}}\ and\ \bibinfo {author} {\bibfnamefont {T.}~\bibnamefont
  {Takayanagi}},\ }\enquote {\bibinfo {title} {Holographic Entanglement Entropy},}\ \
  (\bibinfo  {publisher} {Springer Berlin Heidelberg},\ \bibinfo {address}
  {Berlin, Heidelberg},\ \bibinfo {year} {2017}),\ 
  \Eprint {https://arxiv.org/abs/1609.01287} {arXiv:1609.01287v2 [hep-th]}\BibitemShut 
  {NoStop}%
\bibitem [{\citenamefont {D'Antoni}\ and\ \citenamefont
  {Hollands}(2006)}]{AntoniHollands_CMP2006}%
  \BibitemOpen
  \bibfield  {author} {\bibinfo {author} {\bibfnamefont {C.}~\bibnamefont
  {D'Antoni}}\ and\ \bibinfo {author} {\bibfnamefont {S.}~\bibnamefont
  {Hollands}},\ }{\enquote {\bibinfo {title} {Nuclearity, local quasiequivalence and split property for {D}irac quantum fields in curved spacetime},}}\
  \href {\doibase 10.1007/s00220-005-1398-2} {\bibfield
  {journal} {\bibinfo  {journal} {Comm. Math. Phys.}\ }\textbf {\bibinfo
  {volume} {261}},\ \bibinfo {pages} {133} (\bibinfo {year} {2006})},\ \Eprint
  {http://arxiv.org/abs/0106028v3} {arXiv:0106028v3 [math-ph]}\BibitemShut
  {NoStop}%
\bibitem [{\citenamefont {B{\"a}r}\ and\ \citenamefont
  {Ginoux}(2012)}]{baer_springer2012}%
  \BibitemOpen
  \bibfield  {author} {\bibinfo {author} {\bibfnamefont {C.}~\bibnamefont
  {B{\"a}r}}\ and\ \bibinfo {author} {\bibfnamefont {N.}~\bibnamefont
  {Ginoux}},\ }\enquote {\bibinfo {title} {Global differential geometry},}\ \
  (\bibinfo  {publisher} {Springer Berlin Heidelberg},\ \bibinfo {address}
  {Berlin, Heidelberg},\ \bibinfo {year} {2012})\ Chap.\ \bibinfo {chapter}
  {Classical and Quantum Fields on Lorentzian Manifolds}, pp.\ \bibinfo {pages}
  {359--400},\ \Eprint {http://arxiv.org/abs/1104.1158v2} {arXiv:1104.1158v2
  [math-ph]}\BibitemShut 
  {NoStop}%
\bibitem [{\citenamefont {Kugo}\ and\ \citenamefont
  {Townsend}(1983)}]{KugoTownsend_NuclPhysB1983}%
  \BibitemOpen
  \bibfield  {author} {\bibinfo {author} {\bibfnamefont {T.}~\bibnamefont
  {Kugo}}\ and\ \bibinfo {author} {\bibfnamefont {P.}~\bibnamefont
  {Townsend}},\ }{\enquote {\bibinfo {title} {Supersymmetry and the division algebras},}}\
  \href {\doibase
  http://dx.doi.org/10.1016/0550-3213(83)90584-9} {\bibfield  {journal}
  {\bibinfo  {journal} {Nucl. Phys. B}\ }\textbf {\bibinfo {volume} {221}},\
  \bibinfo {pages} {357 } (\bibinfo {year} {1983})}\BibitemShut 
  {NoStop}%
\bibitem [{\citenamefont {Trautman}(2008)}]{Trautman_JGP2008}%
  \BibitemOpen
  \bibfield  {author} {\bibinfo {author} {\bibfnamefont {A.}~\bibnamefont
  {Trautman}},\ }{\enquote {\bibinfo {title} {Connections and the {D}irac operator on spinor bundles},}}\
  \href {\doibase
  http://dx.doi.org/10.1016/j.geomphys.2007.11.001} {\bibfield  {journal}
  {\bibinfo  {journal} {J. Geom. Phys.}\ }\textbf {\bibinfo {volume} {58}},\
  \bibinfo {pages} {238 } (\bibinfo {year} {2008})}\BibitemShut 
  {NoStop}%
\bibitem [{\citenamefont {Araki}(1970)}]{Araki_RIMS1970}%
  \BibitemOpen
  \bibfield  {author} {\bibinfo {author} {\bibfnamefont {H.}~\bibnamefont
  {Araki}},\ }{\enquote {\bibinfo {title} {On quasifree states of {C}{A}{R} and {B}ogoliubov automorphisms},}}\
  \href
  {http://www.ems-ph.org/journals/show_issue.php?issn=0034-5318&vol=6&iss=3}
  {\bibfield  {journal} {\bibinfo  {journal} {Publ. Res. Inst. Math. Sci.}\
  }\textbf {\bibinfo {volume} {6}},\ \bibinfo {pages} {385 } (\bibinfo {year}
  {1970})}\BibitemShut 
  {NoStop}%
\bibitem [{\citenamefont {Balslev}\ and\ \citenamefont
  {Verbeure}(1968)}]{BalslevVerbeure_CMP1968}%
  \BibitemOpen
  \bibfield  {author} {\bibinfo {author} {\bibfnamefont {E.}~\bibnamefont
  {Balslev}}\ and\ \bibinfo {author} {\bibfnamefont {A.}~\bibnamefont
  {Verbeure}},\ }{\enquote {\bibinfo {title} {States on {C}lifford algebras},}}\
  \href {http://projecteuclid.org/euclid.cmp/1103840329}
  {\bibfield  {journal} {\bibinfo  {journal} {Comm. Math. Phys.}\ }\textbf
  {\bibinfo {volume} {7}},\ \bibinfo {pages} {55} (\bibinfo {year}
  {1968})}\BibitemShut 
  {NoStop}%
\bibitem [{\citenamefont {Balslev}\ \emph {et~al.}(1968)\citenamefont
  {Balslev}, \citenamefont {Manuceau},\ and\ \citenamefont
  {Verbeure}}]{Balslev_CMP1968}%
  \BibitemOpen
  \bibfield  {author} {\bibinfo {author} {\bibfnamefont {E.}~\bibnamefont
  {Balslev}}, \bibinfo {author} {\bibfnamefont {J.}~\bibnamefont {Manuceau}}, \
  and\ \bibinfo {author} {\bibfnamefont {A.}~\bibnamefont {Verbeure}},\ }{\enquote {\bibinfo {title} {Representations of anticommutation relations and {B}ogolioubov transformations},}}\
  \href
  {http://projecteuclid.org/euclid.cmp/1103840645} {\bibfield  {journal}
  {\bibinfo  {journal} {Comm. Math. Phys.}\ }\textbf {\bibinfo {volume} {8}},\
  \bibinfo {pages} {315} (\bibinfo {year} {1968})}\BibitemShut 
  {NoStop}%
  \bibitem [{\citenamefont {Chernoff}(1973)}]{Chernoff_JFA1973}%
  \BibitemOpen
  \bibfield  {author} {\bibinfo {author} {\bibfnamefont {P.~R.}\ \bibnamefont
  {Chernoff}},\ }{\enquote {\bibinfo {title} {Essential self-adjointness of powers of generators of hyperbolic equations},}}\
  \href {\doibase
  http://dx.doi.org/10.1016/0022-1236(73)90003-7} {\bibfield  {journal}
  {\bibinfo  {journal} {J. Func. Anal.}\ }\textbf {\bibinfo {volume} {12}},\
  \bibinfo {pages} {401 } (\bibinfo {year} {1973})}\BibitemShut 
  {NoStop}%
\bibitem [{\citenamefont {Wolf}(1973)}]{Wolf_Indiana1973}%
  \BibitemOpen
  \bibfield  {author} {\bibinfo {author} {\bibfnamefont {J.~A.}\ \bibnamefont
  {Wolf}},\ }{\enquote {\bibinfo {title} {Essential self adjointness for the {D}irac operator and its square},}}\
  \href {https://www.iumj.indiana.edu/IUMJ/FULLTEXT/1973/22/22051}
  {\bibfield  {journal} {\bibinfo  {journal} {Indiana Univ. Math. J.}\ }\textbf
  {\bibinfo {volume} {22}},\ \bibinfo {pages} {611} (\bibinfo {year}
  {1973})}\BibitemShut 
  {NoStop}%
\bibitem [{\citenamefont {D'Antoni}\ and\ \citenamefont
  {Longo}(1983)}]{Antoni_Longo_JFA1983}%
  \BibitemOpen
  \bibfield  {author} {\bibinfo {author} {\bibfnamefont {C.}~\bibnamefont
  {D'Antoni}}\ and\ \bibinfo {author} {\bibfnamefont {R.}~\bibnamefont
  {Longo}},\ }{\enquote {\bibinfo {title} {Interpolation by type I factors and the flip automorphism},}}\
  \href {\doibase http://dx.doi.org/10.1016/0022-1236(83)90018-6}
  {\bibfield  {journal} {\bibinfo  {journal} {J. Func. Anal.}\ }\textbf
  {\bibinfo {volume} {51}},\ \bibinfo {pages} {361 } (\bibinfo {year}
  {1983})}\BibitemShut 
  {NoStop}%
\bibitem [{\citenamefont {Doplicher}\ and\ \citenamefont
  {Longo}(1984)}]{DoplicherLongo_InventMath1984}%
  \BibitemOpen
  \bibfield  {author} {\bibinfo {author} {\bibfnamefont {S.}~\bibnamefont
  {Doplicher}}\ and\ \bibinfo {author} {\bibfnamefont {R.}~\bibnamefont
  {Longo}},\ }{\enquote {\bibinfo {title} {Standard and split inclusions of von {N}eumann algebras},}}\
  \href {\doibase 10.1007/BF01388641} {\bibfield  {journal}
  {\bibinfo  {journal} {Invent. math.}\ }\textbf {\bibinfo {volume} {75}},\
  \bibinfo {pages} {493} (\bibinfo {year} {1984})}\BibitemShut 
  {NoStop}%
\bibitem [{\citenamefont {Narnhofer}(2002)}]{Narnhofer_RepMathPhys2002}%
  \BibitemOpen
  \bibfield  {author} {\bibinfo {author} {\bibfnamefont {H.}~\bibnamefont
  {Narnhofer}},\ }{\enquote {\bibinfo {title} {Entanglement, split and nuclearity in quantum field theory},}}\
  \href {\doibase
  https://doi.org/10.1016/S0034-4877(02)80048-9} {\bibfield  {journal}
  {\bibinfo  {journal} {Rep. Math. Phys.}\ }\textbf {\bibinfo {volume} {50}},\
  \bibinfo {pages} {111 } (\bibinfo {year} {2002})}\BibitemShut 
  {NoStop}%
\bibitem [{\citenamefont {Araki}(1976)}]{Araki_RIMS1976}%
  \BibitemOpen
  \bibfield  {author} {\bibinfo {author} {\bibfnamefont {H.}~\bibnamefont
  {Araki}},\ }{\enquote {\bibinfo {title} {Relative entropy of states of von {N}eumann algebras},}}\
  \href
  {http://www.ems-ph.org/journals/show_abstract.php?issn=0034-5318&vol=11&iss=3&rank=9}
  {\bibfield  {journal} {\bibinfo  {journal} {Publ. Res. Inst. Math. Sci.}\
  }\textbf {\bibinfo {volume} {11}},\ \bibinfo {pages} {809 } (\bibinfo {year}
  {1976})}\BibitemShut 
  {NoStop}%
\bibitem [{\citenamefont {Araki}(1977)}]{Araki_RIMS1977}%
  \BibitemOpen
  \bibfield  {author} {\bibinfo {author} {\bibfnamefont {H.}~\bibnamefont
  {Araki}},\ }{\enquote {\bibinfo {title} {Relative entropy of states of von {N}eumann algebras {I}{I}},}}\
  \href
  {http://www.ems-ph.org/journals/show_abstract.php?issn=0034-5318&vol=13&iss=1&rank=8}
  {\bibfield  {journal} {\bibinfo  {journal} {Publ. Res. Inst. Math. Sci.}\
  }\textbf {\bibinfo {volume} {13}},\ \bibinfo {pages} {173 } (\bibinfo {year}
  {1977})}\BibitemShut 
  {NoStop}%
\bibitem [{\citenamefont {Kadison}\ and\ \citenamefont
  {Ringrose}(1997)}]{KadisonRingrose_II_AMS1997}%
  \BibitemOpen
  \bibfield  {author} {\bibinfo {author} {\bibfnamefont {R.}~\bibnamefont
  {Kadison}}\ and\ \bibinfo {author} {\bibfnamefont {J.}~\bibnamefont
  {Ringrose}},\ }\href {https://books.google.de/books?id=h5bMkZTnowAC} {\emph
  {\bibinfo {title} {Fundamentals of the Theory of Operator Algebras - II:
  Advanced theory}}},\ \bibinfo {series} {Graduate Studies in Mathematics},
  Vol.~\bibinfo {volume} {16}\ (\bibinfo  {publisher} {American Mathematical
  Society},\ \bibinfo {address} {USA},\ \bibinfo {year} {1997})\BibitemShut
  {NoStop}%
\bibitem [{\citenamefont {Foit}(1983)}]{Foit_RIMS1983}%
  \BibitemOpen
  \bibfield  {author} {\bibinfo {author} {\bibfnamefont {J.~J.}\ \bibnamefont
  {Foit}},\ }{\enquote {\bibinfo {title} {Abstract twisted duality for quantum free {F}ermi fields},}}\
  \href {\doibase 10.2977/prims/1195182448} {\bibfield  {journal}
  {\bibinfo  {journal} {Publ. RIMS}\ }\textbf {\bibinfo {volume} {19}},\
  \bibinfo {pages} {729} (\bibinfo {year} {1983})}\BibitemShut 
  {NoStop}%
\bibitem [{\citenamefont {Baumgärtel}\ \emph {et~al.}(2002)\citenamefont
  {Baumgärtel}, \citenamefont {Jurke},\ and\ \citenamefont
  {Lledó}}]{Baumgartel_JMP2002}%
  \BibitemOpen
  \bibfield  {author} {\bibinfo {author} {\bibfnamefont {H.}~\bibnamefont
  {Baumgärtel}}, \bibinfo {author} {\bibfnamefont {M.}~\bibnamefont {Jurke}},
  \ and\ \bibinfo {author} {\bibfnamefont {F.}~\bibnamefont {Lledó}},\ }{\enquote {\bibinfo {title} {Twisted duality of the {C}{A}{R}-algebra},}}\
  \href
  {\doibase 10.1063/1.1483376} {\bibfield  {journal} {\bibinfo  {journal} {J.
  Math. Phys.}\ }\textbf {\bibinfo {volume} {43}},\ \bibinfo {pages} {4158}
  (\bibinfo {year} {2002})},\ \Eprint {http://arxiv.org/abs/0204029v1}
  {arXiv:0204029v1 [math-ph]} \BibitemShut 
  {NoStop}%
\bibitem [{\citenamefont {Longo}()}]{Longo}%
  \BibitemOpen
  \bibfield  {author} {\bibinfo {author} {\bibfnamefont {R.}~\bibnamefont
  {Longo}},\ } \bibinfo {note} {Preliminary lecture notes on Conformal nets} \href {https://www.mat.uniroma2.it/~longo/lecture-notes.html}
  {https://www.mat.uniroma2.it/~longo/lecture-notes.html}\BibitemShut
  {NoStop}%
\bibitem [{\citenamefont {Strohmaier}(2000)}]{Strohmaier_CMP2000}%
  \BibitemOpen
  \bibfield  {author} {\bibinfo {author} {\bibfnamefont {A.}~\bibnamefont
  {Strohmaier}},\ }{\enquote {\bibinfo {title} {The {R}eeh--{S}chlieder property for quantum fields on stationary spacetimes},}}\
  \href {\doibase 10.1007/s002200000299} {\bibfield  {journal}
  {\bibinfo  {journal} {Comm. Math. Phys.}\ }\textbf {\bibinfo {volume}
  {215}},\ \bibinfo {pages} {105} (\bibinfo {year} {2000})},\ \Eprint
  {http://arxiv.org/abs/0002054v2} {arXiv:0002054v2 [math-ph]}\BibitemShut
  {NoStop}%
\bibitem [{\citenamefont {Lechner}\ and\ \citenamefont
  {Sanders}(2016)}]{axioms5_5}%
  \BibitemOpen
  \bibfield  {author} {\bibinfo {author} {\bibfnamefont {G.}~\bibnamefont
  {Lechner}}\ and\ \bibinfo {author} {\bibfnamefont {K.}~\bibnamefont
  {Sanders}},\ }{\enquote {\bibinfo {title} {Modular nuclearity: A generally covariant perspective},}}\
  \href {\doibase 10.3390/axioms5010005} {\bibfield  {journal}
  {\bibinfo  {journal} {Axioms}\ }\textbf {\bibinfo {volume} {5}},\ \bibinfo
  {pages} {5} (\bibinfo {year} {2016})},\ \Eprint
  {http://arxiv.org/abs/1511.09027v1} {arXiv:1511.09027v1 [math-ph]}\BibitemShut 
  {NoStop}%
\bibitem [{\citenamefont {Shimizu}(1985)}]{Shimizu_PTP1985}%
  \BibitemOpen
  \bibfield  {author} {\bibinfo {author} {\bibfnamefont {K.}~\bibnamefont
  {Shimizu}},\ }{\enquote {\bibinfo {title} {C, {P} and {T} transformations in higher dimensions},}}\
  \href {\doibase 10.1143/PTP.74.610} {\bibfield  {journal}
  {\bibinfo  {journal} {Prog. Theor. Phys.}\ }\textbf {\bibinfo {volume}
  {74}},\ \bibinfo {pages} {610} (\bibinfo {year} {1985})}\BibitemShut
  {NoStop}%
\bibitem [{\citenamefont {Winkler}\ and\ \citenamefont
  {Z\"{u}licke}(2015)}]{WinklerZulicke_2015}%
  \BibitemOpen
  \bibfield  {author} {\bibinfo {author} {\bibfnamefont {R.}~\bibnamefont
  {Winkler}}\ and\ \bibinfo {author} {\bibfnamefont {U.}~\bibnamefont
  {Z\"{u}licke}},\ }{\enquote {\bibinfo {title} {Discrete symmetries of low-dimensional {D}irac models: A selective reivew with a focus on condensed-matter realization},}}\
  \href {\doibase 10.1017/S1446181115000115} {\bibfield
  {journal} {\bibinfo  {journal} {ANZIAM J.}\ }\textbf {\bibinfo {volume}
  {57}},\ \bibinfo {pages} {3–17} (\bibinfo {year} {2015})},\ \Eprint
  {http://arxiv.org/abs/1206.0355v2} {arXiv:1206.0355v2 [math-ph]}\BibitemShut
  {NoStop}%
\bibitem [{\citenamefont {Freedman}\ and\ \citenamefont
  {Proeyen}(2012)}]{FreedmanProeyen_CUP2012}%
  \BibitemOpen
  \bibfield  {author} {\bibinfo {author} {\bibfnamefont {D.~Z.}\ \bibnamefont
  {Freedman}}\ and\ \bibinfo {author} {\bibfnamefont {A.~V.}\ \bibnamefont
  {Proeyen}},\ }\href {http://www.cambridge.org/9780521194013} {\emph {\bibinfo
  {title} {Supergravity}}},\ Theoretical Physics and Mathematical Physics\
  (\bibinfo  {publisher} {Cambridge University Press},\ \bibinfo {address}
  {UK},\ \bibinfo {year} {2012})\BibitemShut 
  {NoStop}%
\bibitem [{\citenamefont {Schr{\"o}dinger}(1932)}]{Schrodinger_1932}%
  \BibitemOpen
  \bibfield  {author} {\bibinfo {author} {\bibfnamefont {E.}~\bibnamefont
  {Schr{\"o}dinger}},\ }{\enquote {\bibinfo {title} {Diracsches Elektron im Schwerefeld {I}},}}\
  \href@noop {} {\bibfield  {journal} {\bibinfo
  {journal} {Sit. Preu. Aka. Wiss. Phil.-Hist.}\ }\textbf {\bibinfo {volume}
  {11}},\ \bibinfo {pages} {105 } (\bibinfo {year} {1932})}\BibitemShut
  {NoStop}%
\bibitem [{\citenamefont {Lichnerowicz}(1963)}]{Lichnerowicz_1963}%
  \BibitemOpen
  \bibfield  {author} {\bibinfo {author} {\bibfnamefont {A.}~\bibnamefont
  {Lichnerowicz}},\ }{\enquote {\bibinfo {title} {Spineurs harmoniques},}}\
  \href@noop {} {\bibfield  {journal} {\bibinfo  {journal}
  {C. R. Acad. Sci. Paris}\ }\textbf {\bibinfo {volume} {257}},\ \bibinfo
  {pages} {7 } (\bibinfo {year} {1963})}\BibitemShut {NoStop}%
\bibitem [{\citenamefont {Cheeger}\ \emph {et~al.}(1982)\citenamefont
  {Cheeger}, \citenamefont {Gromov},\ and\ \citenamefont
  {Taylor}}]{Cheeger_JDG1982}%
  \BibitemOpen
  \bibfield  {author} {\bibinfo {author} {\bibfnamefont {J.}~\bibnamefont
  {Cheeger}}, \bibinfo {author} {\bibfnamefont {M.}~\bibnamefont {Gromov}}, \
  and\ \bibinfo {author} {\bibfnamefont {M.}~\bibnamefont {Taylor}},\ }{\enquote {\bibinfo {title} {Finite propagation speed, kernel estimates for functions of the {L}aplace operator, and the geometry of complete {R}iemannian manifolds},}}\
  \href
  {http://projecteuclid.org/euclid.jdg/1214436699} {\bibfield  {journal}
  {\bibinfo  {journal} {J. Differential Geom.}\ }\textbf {\bibinfo {volume}
  {17}},\ \bibinfo {pages} {15} (\bibinfo {year} {1982})}\BibitemShut 
  {NoStop}%
\end{thebibliography}
\end{document}